\documentclass[aps, pra, twocolumn, amsmath, amssymb, superscriptaddress, nofootinbib]{revtex4-1}

\makeatletter
\def\p@subsection{}
\def\p@subsubsection{}
\makeatother

\usepackage[utf8]{inputenc}
\usepackage{graphicx}
\usepackage{units}
\usepackage{color}
\usepackage[pdftex, colorlinks=true, linkcolor=myblue, citecolor=myblue,
urlcolor=myblue]{hyperref}
\usepackage{times}
\usepackage{comment}
\usepackage{graphicx}
\usepackage{amsmath, calc}
\usepackage{mathrsfs}
\usepackage{amsfonts}
\usepackage{amssymb}
\usepackage{bm}
\usepackage{bbm}
\usepackage{color}
\usepackage{array}
\usepackage{units}

\definecolor{myblue}{rgb}{0,0,1}
\definecolor{myred}{rgb}{1,0,0}

\newcommand{\ket}[1]{|#1\rangle}

\newcommand{\bratwo}[1]{\langle \langle #1||}
\newcommand{\kettwo}[1]{||#1\rangle \rangle}

\begin{document}
\title{Resonance fluorescence of two asymmetrically pumped and coupled two-level systems}

\author{C.~A.~Downing}
\email{c.a.downing@exeter.ac.uk} 
\affiliation{Department of Physics and Astronomy, University of Exeter, Exeter EX4 4QL, United Kingdom}

\author{E.~del Valle}
\affiliation{Departamento de F\'{i}sica T\'{e}orica de la Materia Condensada and Condensed Matter Physics Center (IFIMAC), Universidad Aut\'{o}noma de Madrid, E-28049 Madrid, Spain}
\affiliation{Institute for Advanced Study, Technical University of Munich,
Lichtenbergstrasse 2a, D-85748 Garching, Germany}

\author{A.~I.~Fern\'{a}ndez-Dom\'{i}nguez}
\affiliation{Departamento de F\'{i}sica T\'{e}orica de la Materia Condensada and Condensed Matter Physics Center (IFIMAC), Universidad Aut\'{o}noma de Madrid, E-28049 Madrid, Spain}

\date{\today}

\begin{abstract}
\noindent
\\
We study a driven-dissipative duo of two-level systems in an open quantum systems approach, modelling a pair of atoms or (more generally) meta-atoms. Allowing for complex-valued couplings in the setup, which are of both a coherent and incoherent character, gives rise to a diverse coupling landscape. We consider several points on this landscape, for example where the coupling between the two coupled two-level systems is dominated by coherent, incoherent, unsymmetrical and even unidirectional interactions. Traversing the coupling terrain leads to remarkable features in the populations of the pair, correlations and optical spectra. Most notably, the famous Mollow triplet spectrum for a single atom may be superseded for a pair by a Mollow quintuplet (or even by a spectral singlet) and the setup allows for population trapping to arise, all depending upon the precise nature of the coupling between the two-level systems.
\\
\end{abstract}


\maketitle



\section{\label{intro}Introduction}

The theory of resonance fluorescence, which describes the emission of an atom driven resonantly by an external field, has fascinated quantum opticians since the 1960's~\cite{Rautian1962, Newstein1968, Mollow1969, Stroud1971, Kimble1976, Kimble1977b}. Strikingly, the resulting resonance fluorescence spectrum is a so-called Mollow triplet: a central peak at resonance, with two smaller satellite peaks either side~\cite{Mollow1969}. This captivating structure was first seen experimentally in the 1970's~\cite{Schuda1974, Wu1975, Hartig1976, Kimble1977}, before being later observed in single dye molecules~\cite{Wrigge2008} and semiconductor quantum dots~\cite{Xu2007, Muller2007, Ates2009, Flagg2009, Vamivakas2009, Ulhaq2012}. More recently, artificial atoms in superconducting circuits~\cite{Baur2009, Astafiev2010, Lu2019} and hybrid spin-nanomechanical systems~\cite{Pigeau2015} have been shown to display some remarkable aspects of Mollow physics.

Resonance fluorescence in two-atom~\cite{Agarwal1977, Jahangir1978, Mavroyannis1978, Carmichael1979, Kilin1979, Agarwal1980, Gilmore1980, Kus1981, Griffin1982, Ficek1990, Rudolph1995, Skornia2001, Ge2013, Darsheshdar2021, Vivas2021, Ficek2002} and indeed many-atom~\cite{Senitzky1978, Weihan1986, Ott2013, Jenkins2016} systems was inevitably studied theoretically soon afterwards in order to elucidate the influence of cooperative effects, including the emergence of additional sidebands in the optical spectrum. More latterly, modern experiments with two artificial atoms in superconducting circuits have offered the control and tunability required to study the properties of quantum dimers under coherent excitation~\cite{Loo2013}. 

\begin{figure}[tb]
 \includegraphics[width=1.0\linewidth]{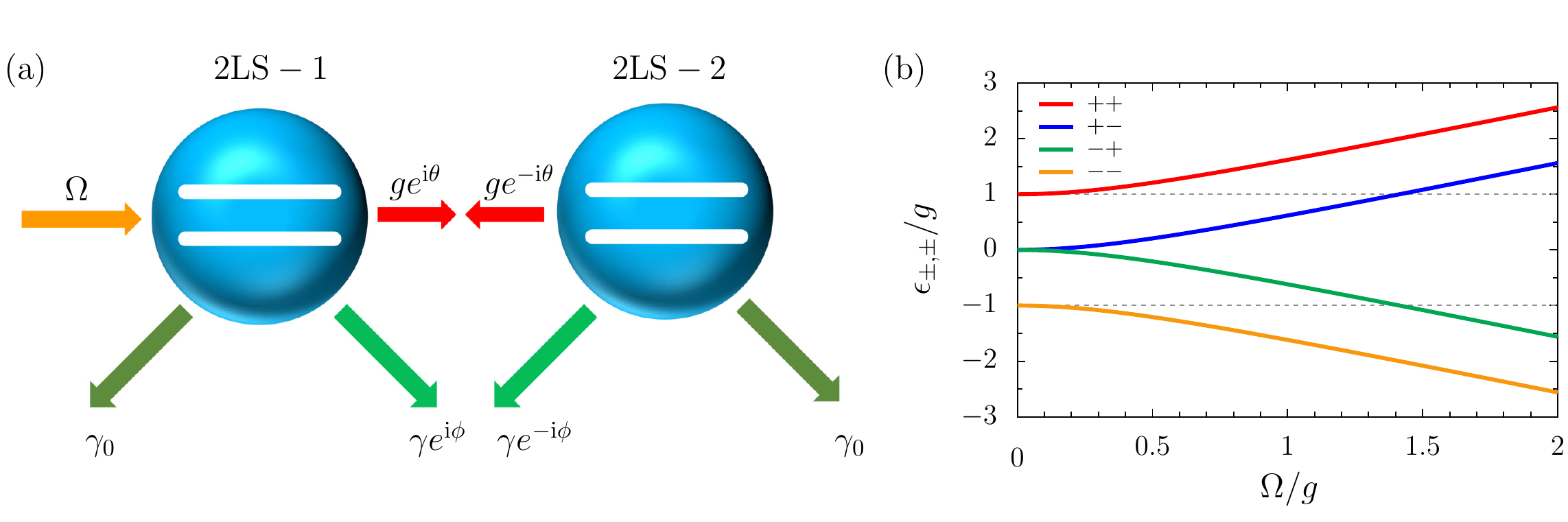}
 \caption{ \textbf{A pair of coupled two-level systems.} Panel (a): A cartoon of the system under investigation: a pair of 2LSs labeled $1$ and $2$, each of transition frequency $\omega_0$ and inverse lifetime $\gamma_0$ (dark green arrows), with both coherent coupling $g \mathrm{e}^{\pm\mathrm{i} \theta}$ (red arrows) and dissipative coupling $\gamma \mathrm{e}^{\pm\mathrm{i} \phi}$ (light green arrows) connecting them [cf. Eq.~\eqref{eq:master}]. The first 2LS is driven coherently (orange arrow) with the amplitude $\Omega$. Panel (b): Energy levels $\epsilon_{\pm, \pm}$ at resonance and in the rotating frame as a function of the drive $\Omega$, both in units of the coherent coupling strength $g$ [cf. Eq.~\eqref{eq:0sdsddssd1}]. Dashed lines: energy levels $\pm g$ in the absence of any driving and in the rotating frame.  }
 \label{sketch}
\end{figure}

Here we investigate theoretically a pair of two-level systems (2LSs) as sketched in Fig.~\ref{sketch}~(a), and in particular the interplay between cooperative resonance fluorescence and the concept of chirality~\cite{Lodahl2017, Andrews2018}. By chirality, we mean to refer to an asymmetry in the coupling between the two 2LSs, which arises from the competition between the considered coherent and incoherent (or dissipative) coupling~\cite{Downing2019, Wang2020, Downing2020}. In an important limiting case, we treat the extreme asymmetry of unidirectional (or one-way) coupling in the pair, where all backaction is excluded by design~\cite{Gardiner1993, Carmichael1993, Metelmann2015}. In this way, we explore the full gamut of Mollow and chiral physics within perhaps the simplest possible coupled system, with a view to building intuition about larger collections of qubits and quantum networks. In particular, chiral quantum networks could transmit information highly efficiently and without information back-flow, while ultracompact chiral devices (acting like circulators and isolators) are necessary to build nanoscale circuits~\cite{Lodahl2017}. 

Our simple model considers the two coupled 2LSs in an open quantum systems approach. We study how the mean populations in the pair, as well as the correlations~\cite{Camilo2015a, Camilo2017,Camilo2018,Wolf2020} and the optical spectra, evolve as one navigates the complex coupling landscape. Most interestingly, the famous Mollow triplet spectrum for a single atom may be superseded for a pair of 2LSs by a range of spectra, from a Mollow quintuplet to a standard Lorentzian singlet, all depending upon the exact nature of the coupling between the pair. We also discover an example of a population trapping effect, where the system is essentially protected from the dissipative environment in a specific part of the coupling landscape. 

The remainder of this paper is assembled in the following manner. We expound the driven-dissipative theory in Sec.~\ref{model}, before focusing on the populations, correlations and spectra of the system in the coherent [Sec.~\ref{cohcoh}], dissipative [Sec.~\ref{dissdiss}], and unidirectional [Sec.~\ref{uniuni}] coupling regimes. Sec.~\ref{conc} contains a discussion of the most important conclusions. Some supporting results for a single 2LS [Appendix~\ref{app:single_2LS}], extra calculational details for the 2LS pair [Appendix~\ref{app:pop}], and a brief survey of asymmetric coupling regime [Appendix~\ref{gengen}] are provided in the three appendices.
\\

\section{\label{model}Model}

Our model is composed of a Hamiltonian contribution which describes the coherent coupling and coherent driving (as introduced in Sec.~\ref{jambon123}) and dissipation which is introduced via a quantum master equation (as defined in Sec.~\ref{jambon456}). The theoretical framework is somewhat analogous to the series of works given by Refs.~\cite{Laussy2009, ValleFermion2009, ValleLaussy2009, delValle2010, delVallePRL2010, ValleLaussy2011} on celebrated models of open quantum systems.
\\


\subsection{\label{jambon123}Hamiltonian}

The Hamiltonian $H$ of a coherently driven pair of coupled 2LSs reads
\begin{equation}
\label{eq:00}
 H = H_0 + H_{\mathrm{c}} + H_{\mathrm{d}},
\end{equation}
where the excitation, coupling and driving Hamiltonians respectively are given by~\cite{Allen1975}
\begin{subequations}
\label{eq:01}
\begin{alignat}{3}
 H_0 &= \omega_0 \left( \sigma_1^{\dagger} \sigma_1 + \sigma_2^{\dagger} \sigma_2 \right),  \label{eq:01aa} \\
 H_{\mathrm{c}} &= g \mathrm{e}^{\mathrm{i} \theta} \sigma_1^{\dagger} \sigma_2 + g \mathrm{e}^{- \mathrm{i} \theta} \sigma_2^{\dagger} \sigma_1, \label{eq:01bb} \\
 H_{\mathrm{d}} &=  \Omega \left( \mathrm{e}^{\mathrm{i} \omega_{\mathrm{d}} t} \sigma_1 + \mathrm{e}^{- \mathrm{i} \omega_{\mathrm{d}} t} \sigma_1^{\dagger} \right), \label{eq:01cc}
 \end{alignat}
\end{subequations}
where $\sigma_n^{\dagger}$ ($\sigma_n$) is the raising (lowering) operator of the $n$-th 2LS, with $n = \{ 1, 2 \}$. These operators satisfy the algebra of two distinguishable systems, with the anticommutator relation $\{ \sigma_n, \sigma_n^{\dagger} \} = 1$, and commutator relations $[ \sigma_n, \sigma_m^{\dagger} ] = [ \sigma_n, \sigma_m ] = 0, n \ne m$. In the excitation Hamiltonian $H_0$, the common resonance frequency of both 2LSs is $\omega_0$, and the four bare states arising from $H_0$ are simply $\{ \ket{0, 0}, \ket{1, 0}, \ket{0, 1}, \ket{1, 1} \}$, which are associated with the ascending eigenfrequencies $\{ 0, \omega_0, \omega_0, 2 \omega_0 \}$. In the coupling Hamiltonian $H_{\mathrm{c}}$, the coherent 2LS-2LS coupling is of magnitude $g \ge 0$ and complex argument $\theta \in [0, 2\pi]$ in order to capture possible interference effects~\cite{Wolf2020}. Within the driving Hamiltonian $H_{\mathrm{d}}$, the driving amplitude is $\Omega$ and its frequency is $\omega_{\mathrm{d}}$. Notably, we have written down Eq.~\eqref{eq:01} in the rotating-wave approximation, which is valid when $g \ll \omega_0$ and $\Omega \ll \omega_0$, such that the counter-rotating terms have been dropped~\cite{Toghill2022}.

Moving Eq.~\eqref{eq:00} into a rotating frame $\tilde{H} = U H U^{\dagger} + \mathrm{i} \left( \partial_{t} U \right) U^{\dagger}$, with the aid of the operator $U = \mathrm{e}^{\mathrm{i} \omega_{\mathrm{d}} t ( \sigma_1^{\dagger} \sigma_1 + \sigma_2^{\dagger} \sigma_2 ) }$, yields the transformed Hamiltonian $\tilde{H}$, as defined by
\begin{align}
\label{eq:02}
 \tilde{H} =& ~\Delta \left( \sigma_1^{\dagger} \sigma_1 + \sigma_2^{\dagger} \sigma_2 \right) + g \left( \mathrm{e}^{\mathrm{i} \theta} \sigma_1^{\dagger} \sigma_2 + \mathrm{e}^{- \mathrm{i} \theta} \sigma_2^{\dagger} \sigma_1 \right) \nonumber \\
  &+ \Omega \left( \sigma_1 + \sigma_1^{\dagger} \right),
\end{align}
where the detuning frequency $\Delta = \omega_0 - \omega_{\mathrm{d}}$. In the basis of bare states $\{ \ket{0, 0}, \ket{1, 0}, \ket{0, 1}, \ket{1, 1} \}$, the $4 \times 4$ matrix representation of the transformed Hamiltonian $\tilde{H}$ of Eq.~\eqref{eq:02} is
\begin{equation}
\label{eqapp:Ham_re232p}
\tilde{H} =
\begin{pmatrix}
  0 &  \Omega &  0 & 0  \\
  \Omega & \Delta & g \mathrm{e}^{\mathrm{i} \theta} & 0 \\
  0 & g \mathrm{e}^{-\mathrm{i} \theta} & \Delta & \Omega \\
  0 & 0 & \Omega & 2 \Delta \\
 \end{pmatrix}.
 \end{equation}
The eigenvalues of Eq.~\eqref{eqapp:Ham_re232p} are the four eigenfrequencies of the system, due its four-dimensional Hilbert space. In the weak driving limit ($\Omega \to 0$), we find the eigenfrequencies $\{ 0, \Delta-g, \Delta+g, 2\Delta \}$, which are associated with the weak driving eigenstates $\{ \ket{\mathrm{G}}, \ket{-}, \ket{+}, \ket{\mathrm{X}} \}$. Here $\ket{\pm} = ( \ket{1, 0} \pm \mathrm{e}^{\mathrm{i} \theta} \ket{0, 1} ) / \sqrt{2} $ are the usual symmetric and antisymmetric intermediate states, while the ground state $\ket{\mathrm{G}} = \ket{0, 0}$ and the doubly-excited state $\ket{\mathrm{X}} = \ket{1, 1}$. In the strong driving limit, with the eigenfrequencies $\{ \omega_{\alpha}, \omega_{\beta}, \omega_{\gamma}, \omega_{\delta} \}$, the inter-level separations increase with increased drivings, and the eigenstates $\{ \kettwo{\alpha}, \kettwo{\beta}, \kettwo{\gamma}, \kettwo{\delta} \}$ become superpositions of all four bare states (throughout this work, we denote the eigenstates in the weak driving regime by $\ket{i}$, and in the strong driving regime by $\kettwo{i}$). In particular, at resonance ($\Delta = 0$) which is the case we mostly consider in this work, the eigenfrequencies of $\tilde{H} $ may be described by the compact expression
\begin{equation}
\label{eq:0sdsddssd1}
\epsilon_{\pm, \pm} = \: \pm \frac{f}{2} \: \pm \frac{g}{2},
\end{equation}
where we have introduced the driving-dependent frequency
\begin{equation}
\label{eq:0sdsddssd12323232}
 f =  \sqrt{  g^2 + 4 \Omega^2 }.
\end{equation}
We plot the energy levels $\epsilon_{\pm, \pm}$ in Fig.~\ref{sketch}~(b) as a function of the driving strength $\Omega$. Notably, the weak driving energy levels $\pm g$ and $0$ (twice) reflect the formed hybridized modes, which tend towards $\pm g/2 \pm \Omega$ with strong driving. The differences between the energy levels of Eq.~\eqref{eq:0sdsddssd1}, that is $f, g, f+g$ and $f-g$, are related to the transition frequencies between the dressed states and are hence important for the optical spectrum of the system, as will be shown later on. The introduction of dissipation necessarily changes the behaviour of the energy levels shown in Fig.~\ref{sketch}~(b) as we now consider.
\\

\begin{figure}[tb]
 \includegraphics[width=1.0\linewidth]{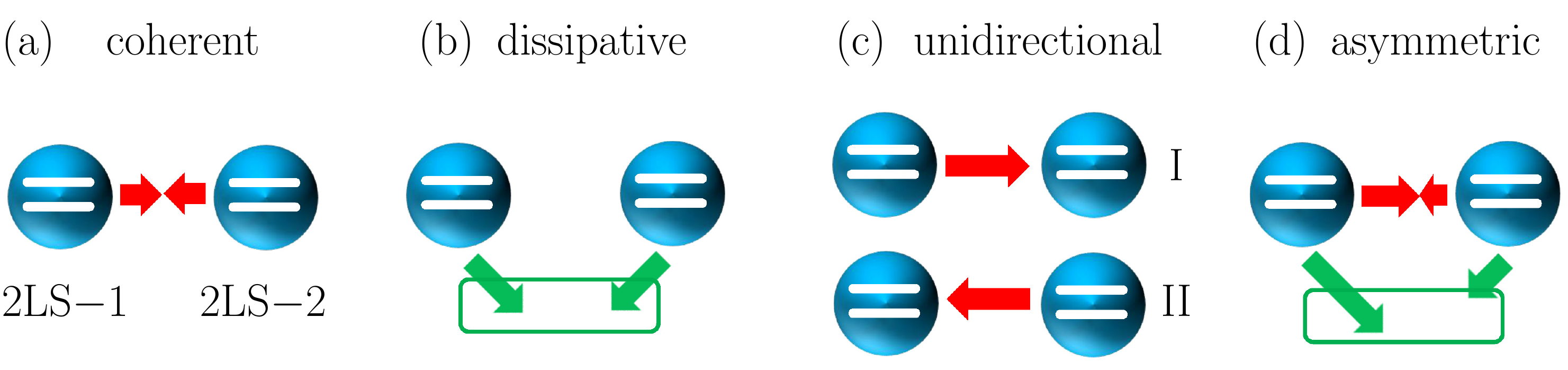}
 \caption{ \textbf{The coupling landscape.} Cartoons of the open quantum system under consideration, showcasing some of most important coupling regimes which form part of the overall coupling landscape. The red arrows are associated with coherent coupling, while the green arrows suggest incoherent (dissipative) coupling, which arises due to the heat bath (green rectangle) which is shared by other 2LSs (marked as $1$ and $2$). Panel (a): coherent coupling regime of Sec.~\ref{cohcoh}. Panel (b): incoherent (dissipative) coupling regime of Sec.~\ref{dissdiss}. Panel (c): unidirectional coupling regime of Sec.~\ref{uniuni}. Panel (d): asymmetric coupling regime of Appendix~\ref{gengen}.}
 \label{regimes}
\end{figure}


\subsection{\label{jambon456}Quantum master equation}

\begin{figure*}[tb]
 \includegraphics[width=\linewidth]{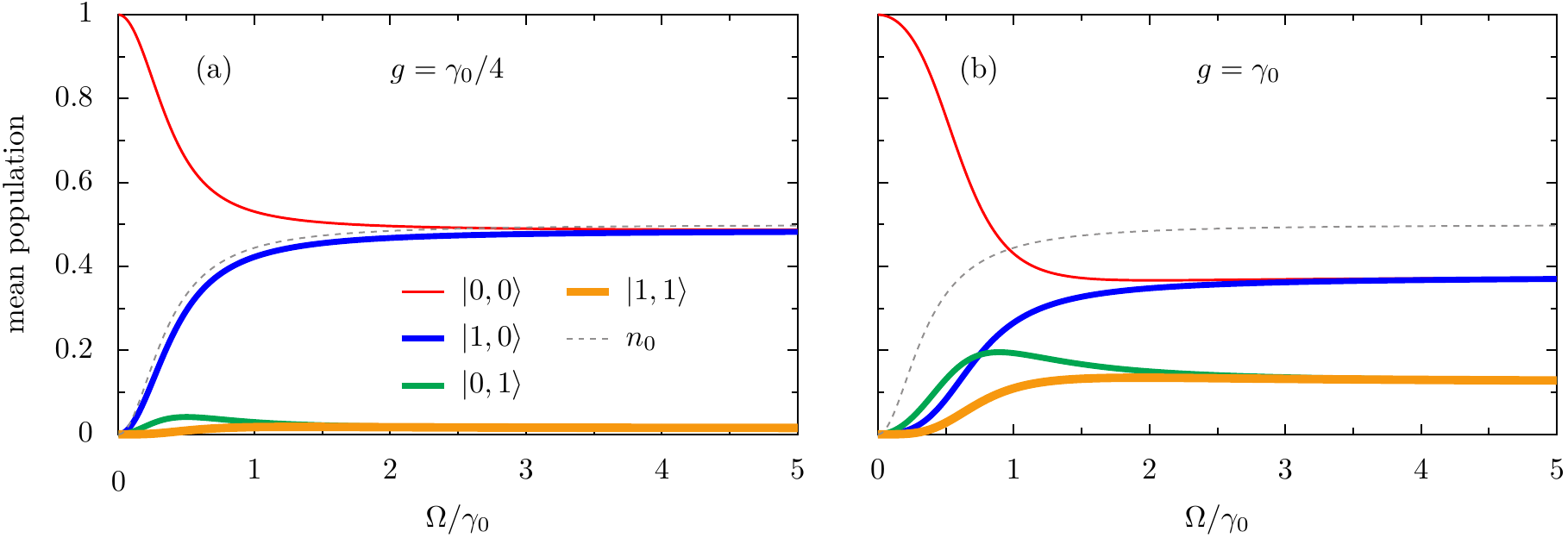}
 \caption{ \textbf{Populations in the coherent coupling regime.} Average populations of the 2LS duo when the coherent coupling dominates, plotted against the drive amplitude $\Omega$ [cf. Eq.~\eqref{eq:statespopgen}]. The single 2LS population is shown as the dashed gray line [cf. Eq.~\eqref{eq:popuionsd343434issi}]. The coherent coupling strength $g = \gamma_0/4$ in panel (a), while $g = \gamma_0$ in panel (b). The legend inside panel (a) marks the result for the state $\ket{n, m}$.}
 \label{popco}
\end{figure*}

Assuming the Born, Markov and rotating wave approximations allows for the quantum master equation of the system's density matrix $\rho$ to be written in the standard Lindblad form~\cite{Breuer2002, Gardiner2004, Gardiner2014}
\begin{align}
\label{eq:master}
 \partial_t \rho = \ &\mathrm{i} [ \rho, \tilde{H} ] \nonumber \\
  &+ \frac{\gamma_{0}}{2} \left( 2 \sigma_1 \rho \sigma_1^{\dagger} -  \sigma_1^{\dagger} \sigma_1 \rho - \rho \sigma_1^{\dagger} \sigma_1 \right) \nonumber \\
  &+ \frac{\gamma_{0}}{2} \left( 2 \sigma_2 \rho \sigma_2^{\dagger} -  \sigma_2^{\dagger} \sigma_2 \rho - \rho \sigma_2^{\dagger} \sigma_2 \right) \nonumber \\
  &+ \frac{\gamma}{2} \mathrm{e}^{\mathrm{i} \phi} \left( 2 \sigma_2 \rho \sigma_1^{\dagger} -  \sigma_1^{\dagger} \sigma_2 \rho - \rho \sigma_1^{\dagger} \sigma_2 \right) \nonumber \\
  &+ \frac{\gamma}{2} \mathrm{e}^{-\mathrm{i} \phi} \left( 2 \sigma_1 \rho \sigma_2^{\dagger} -  \sigma_2^{\dagger} \sigma_1 \rho - \rho \sigma_2^{\dagger} \sigma_1 \right), 
\end{align}
where the rotated Hamiltonian $\tilde{H}$ is given by Eq.~\eqref{eq:02}, and $\gamma_0$ is the decay rate of each individual 2LS. The dissipative (or incoherent) coupling is of magnitude $0 \le \gamma \le \gamma_0$ and phase $\phi \in [0, 2\pi]$, in a similar manner to the complex coherent coupling $g \mathrm{e}^{\mathrm{i} \theta}$ introduced in Eq.~\eqref{eq:01bb}. In any observables resulting from Eq.~\eqref{eq:master}, it is the relative phase $\theta - \phi$ between the coherent and incoherent couplings which is important, and not the absolute phases $\theta$ or $\phi$ on their own. All parameters entering our open quantum systems model described by Eq.~\eqref{eq:master}, along with Eq.~\eqref{eq:02}, are represented in Fig.~\ref{sketch}~(a).

With regard to specific examples where a relative phase arises between the coherent and incoherent coupling parameters, consider the following. Within the macroscopic quantum electrodynamical theory of a pair of qubits with dipole-dipole interactions~\cite{Dung2002}, the aforementioned coupling parameters ($g \mathrm{e}^{\pm \mathrm{i} \theta}, \gamma \mathrm{e}^{\pm \mathrm{i} \phi}$) are defined in terms of the real and imaginary parts of the classical Green's tensor for the electromagnetic environment and the dipole moments of the qubits. In the setup of Ref.~\cite{Gonzalez2011} where this theory is applied to qubits above a metallic plane sustaining plasmons (see also Ref.~\cite{Downing2019}), circular polarization of the qubits gives rise to complex phases associated with the coupling parameters (so that, for example, changing the relative positions of the qubits allows one to sweep through the coupling landscape). 

The interplay between the coherent and dissipative coupling strengths, via the ratio of magnitudes $g/\gamma$ as well as their relative phase $\theta - \phi$, allows one to navigate a rich coupling landscape. In particular, the quantum master equation of Eq.~\eqref{eq:master} exactly maps onto the celebrated cascaded master equation introduced by Gardiner and Carmichael in the early 1990s~\cite{Gardiner1993, Carmichael1993}, this mapping occurs when 
\begin{subequations}
\label{eq:conditions}
\begin{align}
 \frac{g}{\gamma} &= \frac{1}{2}, \label{eq:conditions2}  \\
 \theta  &=
\begin{cases} \phi + \tfrac{\pi}{2}, \quad  \quad ~~\text{1}~\rightarrow~\text{2}, \quad \quad (\text{situation}~\mathrm{I}) \\ 
\phi + \tfrac{3 \pi}{2}, \quad  \quad \text{1}~\leftarrow~\text{2}, \quad \quad (\text{situation}~\mathrm{II})
\end{cases}. \label{eq:conditions1}
 \end{align}
\end{subequations} 
This formal mapping implies completely unidirectional (or one-way) effective interactions amongst the duo of 2LSs~\cite{Camilo2015a, Downing2019,Downing2020}.

We sketch the four principle coupling regimes under consideration in Fig.~\ref{regimes}, namely: coherent coupling, dissipative coupling, unidirectional coupling [cf. Eq.~\eqref{eq:conditions}], and asymmetric coupling. In the remainder of this work, we investigate how various quantum optical phenomena change with the undulations of the coupling landscape, discussing each regime in turn. The simplest regime, that of coherent coupling, is treated in Sec.~\ref{cohcoh} and lays the groundwork of the most intuitive cooperative Mollow physics. It includes a detailed analysis of the spectral Mollow quintuplet which can arise with stronger couplings. The dissipative coupling regime of Sec.~\ref{dissdiss} is most notable for allowing for an interesting population trapping effect, whereby a state of the system is effectively immune from decay. Finally, the unidirectional coupling regime of Sec.~\ref{uniuni} shows how an ostensibly coupled system can behave like an uncoupled system, due to the interference of the coherent and incoherent couplings. We consider the system to be at resonance ($\Delta = 0$) throughout the remainder of this work.
\\


\section{\label{cohcoh}The regime of coherent coupling}

Here we focus on the purely coherent ($\mathrm{co}$) coupling regime as sketched in Fig.~\ref{regimes}~(a), where the dissipative coupling is entirely absent (that is, the incoherent coupling parameter is set to $\gamma = 0$). We discuss the average 2LS populations (in Sec.~\ref{popcoh}), the correlations (in Sec.~\ref{popcorr}), and the power spectrum (in Sec.~\ref{speccoh}), and in particular how these quantities change as both the coherent coupling strength $g$ and the driving strength $\Omega$ are modulated.

This coupling regime has perhaps the widest applicability to current and the most readily tunable experimental platforms, including for example with semiconductor quantum dots~\cite{Portalupi2016}, solid state spin qubits~\cite{Joas2017}, laser-driven cold atomic clouds~\cite{Ortiz2019}, and photonic ring cavities~\cite{Cui2021}.
\\


\subsection{\label{popcoh}Populations}

In what follows, we compare our results for the coupled pair of 2LSs to the average population $n_0$ of a solitary, uncoupled 2LS in the steady state (please see Appendix~\ref{app:pop} for details)
\begin{equation}
\label{eq:popuionsd343434issi}
 n_0 = \frac{\Omega^2}{2 \Omega^2 + \left( \tfrac{\gamma_0}{2} \right)^2 },
\end{equation}
where the single 2LS is of decay rate $\gamma_0$ and is driven coherently with the amplitude $\Omega$. Notably, population inversion is prevented with purely coherent driving, since $n_0 \to 1/2$ in the limit of strong driving $\Omega \gg \gamma_0$~\cite{Camilo2016, Camilo2016b}. The analogous average populations of full system of the 2LS duo in the coherent coupling regime $\rho_{n, m}^{\mathrm{co}}$, referring to the four bare states $\ket{n, m}$ are (please see Appendix~\ref{app:pop} for the supporting theory)
\begin{widetext}
\begin{subequations}
\label{eq:statespopgen}
\begin{align}
 \rho_{0, 0}^{\mathrm{co}} &= \frac{ \left( 4 g^2 + 9 \gamma_0^2 \right) \left( 8 \gamma_0^4 \Omega^2 +  \left[ 4 g^2 \gamma_0 + \gamma_0^3 \right]^2 \right)
 + 16 \Omega^4 \left( 4 g^4 + 13 g^2 \gamma_0^2 + 11 \gamma_0^4 \right) 
 + 64 \Omega^6 \left( g^2 + 2 \gamma_0^2 \right)  }{ \left( 4 g^2 + 9 \gamma_0^2 \right) \left( 4 g^2 \gamma_0 + \gamma_0^3 \right)^2
 + 4 \gamma_0^2 \Omega^2 \left( 16 g^4 + 48 g^2 \gamma_0^2 + 27 \gamma_0^4 \right)
 + 64 \Omega^4 \left( 4 g^4 + 11 g^2 \gamma_0^2 + 5 \gamma_0^4 \right)
 + 256 \Omega^6 \left( g^2 + \gamma_0^2 \right) }, \label{eq:mike1} \\
 \rho_{1, 0}^{\mathrm{co}} &= \frac{ \left( 4 g^2 + 9 \gamma_0^2 \right) \left( 4 \gamma_0^4 \Omega^2 + 16 \Omega^4 \left[ g^2 + \gamma_0^2 \right] \right) 
 + 64 \Omega^6 \left( g^2 + 2 \gamma_0^2 \right) }{ \left( 4 g^2 + 9 \gamma_0^2 \right) \left( 4 g^2 \gamma_0 + \gamma_0^3 \right)^2
 + 4 \gamma_0^2 \Omega^2 \left( 16 g^4 + 48 g^2 \gamma_0^2 + 27 \gamma_0^4 \right)
 + 64 \Omega^4 \left( 4 g^4 + 11 g^2 \gamma_0^2 + 5 \gamma_0^4 \right)
 + 256 \Omega^6 \left( g^2 + \gamma_0^2 \right) },  \label{eq:mike2} \\
 \rho_{0, 1}^{\mathrm{co}} &= \frac{ 16 g^2 \Omega^2 \left( 9 \gamma_0^4 + 9 \gamma_0^2 \Omega^2 + 4 \Omega^4 + 4 g^2 \left[ \gamma_0^2 + \Omega^2 \right] \right) }{ \left( 4 g^2 + 9 \gamma_0^2 \right) \left( 4 g^2 \gamma_0 + \gamma_0^3 \right)^2
 + 4 \gamma_0^2 \Omega^2 \left( 16 g^4 + 48 g^2 \gamma_0^2 + 27 \gamma_0^4 \right)
 + 64 \Omega^4 \left( 4 g^4 + 11 g^2 \gamma_0^2 + 5 \gamma_0^4 \right)
 + 256 \Omega^6 \left( g^2 + \gamma_0^2 \right) }, \label{eq:mike3} \\
 \rho_{1, 1}^{\mathrm{co}} &= \frac{ 16 g^2 \Omega^4 \left( 4 g^2 + 9 \gamma_0^2 + 4 \Omega^2 \right) }{ \left( 4 g^2 + 9 \gamma_0^2 \right) \left( 4 g^2 \gamma_0 + \gamma_0^3 \right)^2
 + 4 \gamma_0^2 \Omega^2 \left( 16 g^4 + 48 g^2 \gamma_0^2 + 27 \gamma_0^4 \right)
 + 64 \Omega^4 \left( 4 g^4 + 11 g^2 \gamma_0^2 + 5 \gamma_0^4 \right)
 + 256 \Omega^6 \left( g^2 + \gamma_0^2 \right) }. \label{eq:mike4}
\end{align} 
\end{subequations}
\end{widetext}
In the limit of strong driving ($\Omega \gg \gamma_0$) we find the much simpler asymptotic forms of Eq.~\eqref{eq:statespopgen} are
\begin{align}
\label{eq:statespopgenlimit}
 \rho_{0, 0}^{\mathrm{co}} = \rho_{1, 0}^{\mathrm{co}} = \frac{1}{4} \frac{ g^2 +  2 \gamma_0^2 }{g^2 + \gamma_0^2},  \\
 \rho_{0, 1}^{\mathrm{co}} = \rho_{1, 1}^{\mathrm{co}} = \frac{1}{4} \frac{ g^2 }{g^2 + \gamma_0^2},
\end{align}
showcasing an equality in mean population between two pairs of states. Firstly, the ground state $\ket{0, 0}$ and driven state $\ket{1, 0}$ become equal in population, and secondly the undriven state $\ket{0, 1}$ and doubly-excited state $\ket{1, 1}$ become indistinguishable in population.

We plot in Fig.~\ref{popco} the populations of Eq.~\eqref{eq:statespopgen} as a function of the drive amplitude $\Omega$, while the dashed gray line represents the result for a single 2LS [cf. Eq.~\eqref{eq:popuionsd343434issi}]. In panel (a), where the coherent coupling strength $g = \gamma_0/4$, one notices how only the driven 2LS is non-negligibly populated via $\ket{1, 0}$ (medium blue line), being almost equivalent to the single 2LS result. In the limit of strong driving, this mean population reaches parity with the ground state $\ket{0, 0}$ (thin red line) following the limit in Eq.~\eqref{eq:statespopgenlimit}, while the other states have a negligible chance of being populated. In Fig.~\ref{popco}~(b), with the stronger coherent coupling strength $g = \gamma_0$, the nonzero populations of all four distinct states $\ket{n, m}$ are clearly visible for all non-vanishing driving strengths. Notably, the doubly-excited state $\ket{1, 1}$ (thick orange line) plateaus to the mean population of the non-driven singly-excited state $\ket{0, 1}$ (medium green line) with strong driving, in the manner of Eq.~\eqref{eq:statespopgenlimit}. The difference between the driven $\ket{1, 0}$ population (medium blue line) and the single 2LS result (dashed gray line) is most apparent due to the increased coherent coupling $g$, which allows for both 2LSs to become meaningfully populated.
\\


\subsection{\label{popcorr}Correlations}

\begin{figure}[tb]
 \includegraphics[width=1.0\linewidth]{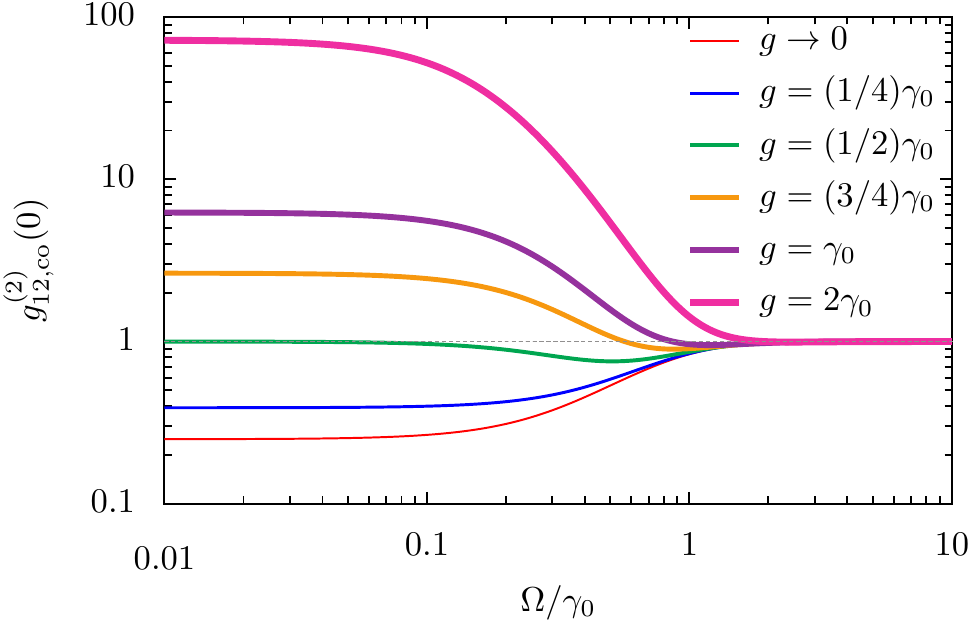}
 \caption{ \textbf{Correlations in the coherent coupling regime.} Cross-correlator $g_{12, \mathrm{co}}^{(2)} (0)$ when there is no time delay, and when the coherent coupling dominates. We plot against the drive amplitude $\Omega$ [cf. Eq.~\eqref{eq:g2co}]. Various coupling strengths $g$ are considered (see the plot legend). Dashed gray line: guide for the eye at unity.}
 \label{corrco}
\end{figure}

\begin{figure*}[tb]
 \includegraphics[width=\linewidth]{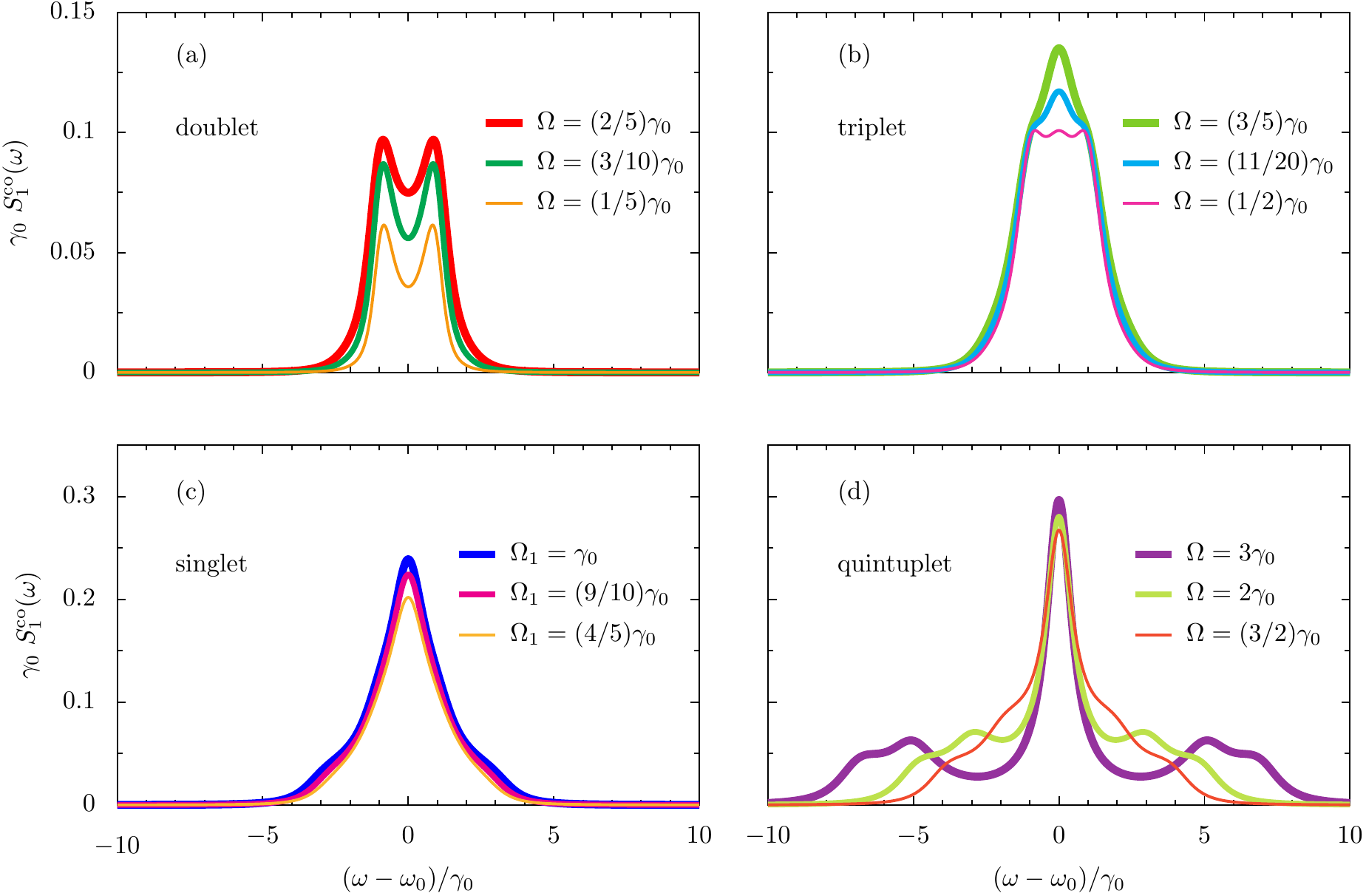}
 \caption{ \textbf{Spectra in the coherent coupling regime.} Optical spectrum $S_1^{\mathrm{co}}(\omega)$ of the first 2LS in the coherent coupling regime, in units of the inverse damping rate $\gamma_0^{-1}$. In each panel, stronger drivings $\Omega$ are associated with thicker lines. The regime of the doublet, with weak driving, is displayed in Panel (a). The triplet regime, with moderate driving, is shown in Panel (b). The singlet regime, with strong driving, is given in Panel (c). Panel (d): the quintuplet regime, with very strong driving. In the figure, the coherent coupling strength $g = \gamma_0$.  }
 \label{specco}
\end{figure*}

In this section, we are concerned with Glauber's (normalized) second-order cross-correlator  $g_{12}^{(2)} (\tau)$ at zero delay (that is, $\tau = 0$), as defined by the expression~\cite{ValleBook2010, Kavokin2007}
\begin{equation}
\label{eq:0galub1}
 g_{12}^{(2)} (0) = \frac{\langle {\sigma_1^{\dagger}} \sigma_1 {\sigma_2^{\dagger}} \sigma_2 \rangle_\mathrm{ss}}{ \langle {\sigma_1^{\dagger}} \sigma_1 \rangle_\mathrm{ss}  \langle {\sigma_2^{\dagger}} \sigma_2 \rangle_\mathrm{ss}},
\end{equation}
as calculated in the steady state ($\mathrm{ss}$). The cross-correlator measures the chance for simultaneous emissions from both 2LSs. When $g_{12}^{(2)} (0) = 1$ the system is uncorrelated, when $g_{12}^{(2)} (0) < 1$ the system is displaying anticorrelations, and when $g_{12}^{(2)} (0) > 1$ the system is exhibiting correlations. The exact expression for the cross-correlator $g_{12, \mathrm{co}}^{(2)} (0)$ in the coherent coupling regime reads (see Appendix~\ref{app:pop} for the theory)
\begin{widetext}
\begin{equation}
\label{eq:g2co}
 g_{12, \mathrm{co}}^{(2)} (0) = 1 + \frac{g^2 \left( 4 g^2 + 3 \gamma_0^2 \right) \left( 4 g^2 + 9 \gamma_0^2 + 4 \Omega^2 \right) }{ 4 g^2 \gamma_0^4 + 9 \gamma_0^6 + 4 \Omega^2 \left(  8 g^4 + 22 g^2 \gamma_0^2 + 9 \gamma_0^4 \right) + 32 \Omega^4 \left( g^2 + \gamma_0^2 \right) } - \frac{ 16 g^4 + 27 \gamma_0^4 - 4 \gamma_0^2 \Omega^2 + 16 g^2 \left( 3 \gamma_0^2 + \Omega^2 \right) }{ 4 \left( 9 \gamma_0^4 + 18 \gamma_0^2 \Omega^2 + 8 \Omega^4 + 4 g^2 \left( \gamma_0^2 + 2 \Omega^2 \right)  \right) }.
\end{equation}
\end{widetext}
In the limits of weak and strong driving respectively, we find from Eq.~\eqref{eq:g2co} the more compact asymptotics 
\begin{align}
\label{eq:g2colim}
 g_{12, \mathrm{co}}^{(2)} (0) &= \frac{ \left( 4 g^2 + \gamma_0^2 \right)^2 }{ 4 \gamma_0^4 },
 \quad 
 &&\Omega \ll \gamma_0,  \\
  \quad 
  g_{12, \mathrm{co}}^{(2)} (0) &= 1,
  \quad 
   &&\Omega \gg \gamma_0. 
\end{align}
Therefore, the system is uncorrelated and $g_{12, \mathrm{co}}^{(2)} (0) \to 1$ with high driving, since the driving $\Omega$ dominates the coherent coupling $g$ and the pair behaves like an uncoupled system. With weak driving, both correlations and anticorrelations are possible. The anticorrelation minimum of $g_{12, \mathrm{co}}^{(2)}(0) \to 1/4$ is seen to arise with vanishing coherent coupling $g \to 0$, while for strong coupling $g \gg \gamma_0$ the correlation maximum follows via a quartic scaling, $g_{12, \mathrm{co}}^{(2)}(0) \simeq 4 (g/\gamma_0)^4$.

We plot the  full cross-correlator $g_{12, \mathrm{co}}^{(2)} (0)$  as a function of the drive amplitude $\Omega$ in Fig.~\ref{corrco}~(a) using Eq.~\eqref{eq:g2co}. In the plot, thicker lines are associated with stronger coherent coupling strengths $g$. Most noticeably, for $\Omega \gg \gamma_0$ all correlations are washed out, which is increasingly the case for smaller coherent coupling $g$ (thinner lines), where one finds $g_{12, \mathrm{co}}^{(2)} (0) \to 1$. At small drivings $\Omega \ll \gamma_0$, the impact of differing coherent coupling strengths $g$ is keenly felt. Approximately, if $g > \gamma_0/2$ (orange, purple and pink thicker lines) correlation behavior $ g_{12, \mathrm{co}}^{(2)} (0) > 1$ is displayed, while for $g < \gamma_0/2$ (blue and red thinner lines) anticorrelations with $ g_{12, \mathrm{co}}^{(2)} (0) < 1$ arises, as follows from the asymptotics of Eq.~\eqref{eq:g2colim}. Hence, in the weak driving limit at least, the system an interesting tunability in $g_{12, \mathrm{co}}^{(2)} (0)$ via the coupling strength $g$.
\\


\subsection{\label{speccoh}Spectrum}

Here we are interested in the optical spectrum in the steady state ($\mathrm{ss}$), as defined via the integral over the two-time correlator $\langle \sigma_1^\dagger (t) \sigma_1 ( t + \tau ) \rangle$~\cite{ValleBook2010, Kavokin2007}
\begin{equation}
\label{eq:speccccc}
 S_1(\omega) = \frac{1}{\pi} \lim_{ t \to  \infty}  \mathrm{Re}~ \frac{  \int_0^\infty \mathrm{e}^{\mathrm{i} \omega \tau} \langle \sigma_1^\dagger (t) \sigma_1 (  \tau + t) \rangle  \mathrm{d}\tau  }{    \langle \sigma_1^\dagger (t) \sigma_1 (t) \rangle},
\end{equation}
that is, the probability density that the first 2LS has an emission at the frequency $\omega$. The theoretical framework for the results is detailed in Appendix~\ref{app:pop}, where we also comment on the Rayleigh (delta) peak which has been neglected throughout the main text. In essence, Eq.~\eqref{eq:speccccc} may be re-written as the summation of Lorentzian and dispersive lineshapes~\cite{delValle2010}
\begin{equation}
\label{eq:pspec2sdfdf343}
S_1 (\omega) = \sum_\zeta \frac{1}{\pi} \frac{\tfrac{\gamma_\zeta}{2}~L_\zeta - \left( \omega - \omega_\zeta \right)~K_\zeta}{ \left( \tfrac{\gamma_\zeta}{2} \right)^2 + \left( \omega - \omega_\zeta \right)^2  }.
\end{equation}
where the spectral peaks $\omega_\zeta$ and the spectral broadenings $\gamma_\zeta$ arise from the possible transitions (indexed with $\zeta$) in the system. The real numbers $L_\zeta$ and $K_\zeta$ are the weighting factors of the Lorentzian and dispersive parts for each spectral component~\cite{delValle2010}. Notably, in the simpler case of the spectrum of a single 2LS all aforementioned spectral parameters $\{ \omega_\zeta, \gamma_\zeta, L_\zeta, K_\zeta \}$ may be calculated analytically, as is shown in Appendix~\ref{app:single_2LS}.

In Fig.~\ref{specco}, we plot the incoherent part of the optical spectrum $S_1^{\mathrm{co}}(\omega)$ of the first 2LS in the coherent and strong coupling regime, where the coherent coupling strength $g = \gamma_0$. In each panel of Fig.~\ref{specco},  stronger drivings $\Omega$ are denoted by thicker colored lines, as indicated by the plot legends. In panel (a), we show the results with weak driving $\Omega \le (2/5) \gamma_0$, where the spectrum is clearly a well-defined doublet due to the emissions from the intermediate states to the ground state (the doubly-excited eigenstate has a negligible population). Panel (b) shows the effect of moderate driving $(2/5) \gamma_0 \le \Omega \le (3/5) \gamma_0$, causing the doublet spectrum to split into a narrow triplet, which is most noticeable for $\Omega = \gamma_0/2$ (thin pink line). However, soon the satellite peaks either side of the main peak, which is always centered at resonance, become almost indistinguishable (thicker cyan and lime lines). In panel (c), with strong driving $(3/5) \gamma_0 \le \Omega \le \gamma_0$, the quasi-singlet regime is eventually reached, although towards the upper end of this regime the mostly hidden satellite peaks are beginning to emerge (thick blue line). Eventually in panel (d), with very strong driving $\Omega \ge \gamma_0$, the Mollow quintuplet regime is unveiled, where either side of the main central peak are doublet satellite peaks. The quintuplet is most visible for the strongest drivings (thick purple line), and is a hallmark of the coherent coupling regime with very strong driving, in direct analogy to the Mollow triplet for a single 2LS.

\begin{figure*}[tb]
 \includegraphics[width=1.0\linewidth]{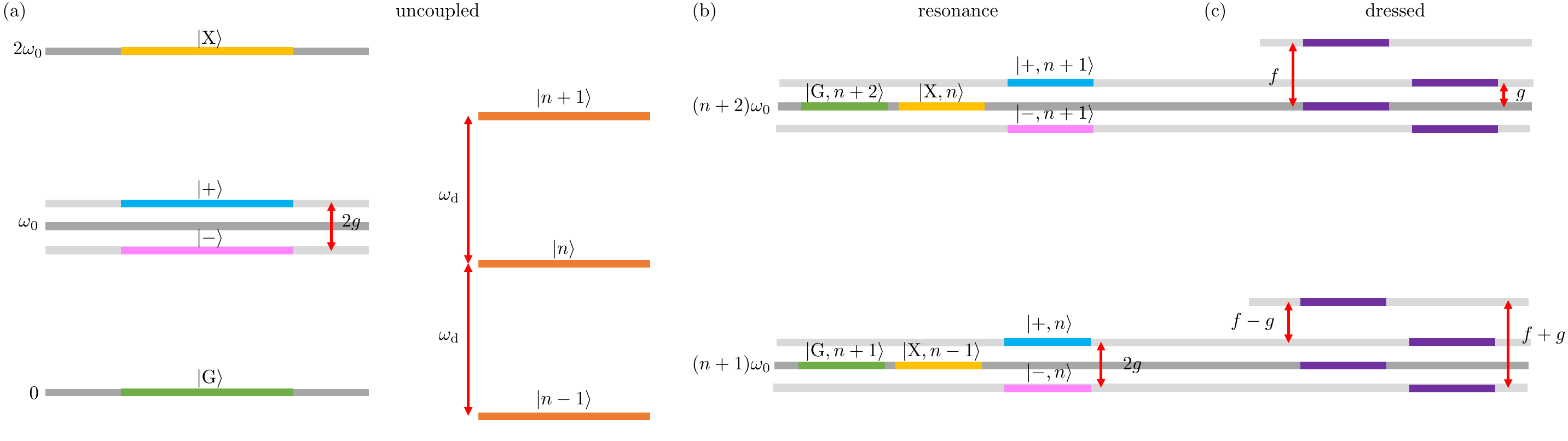}
 \caption{ \textbf{An interpretation of the Mollow quintuplet.} Panel (a): A sketch of the energy ladder giving rise to the Mollow quintuplet. Panel (a): the four states of the coupled two-level systems are represented by coloured bars, including the spitting $2g$ between the intermediate states $\ket{+}$ and $\ket{-}$ (red arrow). The infinite number of states $\ket{n}$ of the quantum harmonic oscillator (modelling the driving laser) are depicted by orange bars, and the inter-level spacing $\omega_{\mathrm{d}}$ is shown (red arrows). Panel (b): at resonance ($\omega_0 = \omega_{\mathrm{d}}$) and in the rotating frame, the combined system of two coupled 2LSs and driving laser displays doubly-degenerate energy levels. For example, at $(n+2)\omega_0$ the degeneracy arises via the states $| \mathrm{G}, n+2 \rangle$ (green bar) and $|  \mathrm{X}, n \rangle$ (yellow bar). The intermediate states $| +, n+1 \rangle$ (cyan bar) and $| -, n+1 \rangle$ (pink bar) lie either side of this degeneracy. Panel (c): The 2LS-driving coupling splits the doubly-degenerate levels leading to a doublet spaced by $f = \sqrt{g^2+4\Omega^2}$. There are then several decay channels from the dressed states (purple bars) and the five transition frequencies $\omega_0$, $\omega_0 \pm (f + g)$ and $\omega_0 \pm (f - g)$ correspond to the five distinct frequencies of the Mollow quintuplet.}
 \label{expl}
\end{figure*}

As discussed in Appendix~\ref{app:single_2LS}, a strongly laser-driven single 2LS is associated with four transitions (two of which are degenerate) leading to the possibility of a triplet spectral structure [cf. Fig.~\ref{single}~(b) for the Mollow triplet and Fig.~\ref{tran} for its explanation]. This interpretation may be extended to a pair of coupled 2LSs and its associated quintuplet structure, as illustrated in Fig.~\ref{expl}. Let us consider the four states $\ket{\mathrm{G}}$, $\ket{-}$, $\ket{+}$ and $\ket{\mathrm{X}}$ arising from the undriven Hamiltonian $H_0 + H_{\mathrm{c}}$ [cf. Eq.~\eqref{eq:01}], and residing at $0$, $\omega_0-g$, $\omega_0+g$ and $2\omega_0$ respectively. We also include the quantum harmonic oscillator states $\ket{n}$ representing the laser driving the system, which are separated in frequency by $\omega_{\mathrm{d}}$ as sketched in Fig.~\ref{expl}~(a). At resonance ($\omega_0 = \omega_{\mathrm{d}}$) and in the rotating frame, the combined system of coupled 2LSs and driving laser displays doubly-degenerate energy levels at $(n+2)\omega_0$, through the states $| \mathrm{G}, n+2 \rangle$ (green bar) and $|  \mathrm{X}, n \rangle$ (yellow bar). The superposition states $| +, n+1 \rangle$ (cyan bar) and $| -, n+1 \rangle$ (pink bar) lie either side of this degeneracy, split by $2g$ (red arrow), as shown in the left-hand-side of Fig.~\ref{expl}~(b). The 2LS-driving coupling then splits the doubly-degenerate levels leading to a doublet spaced by the frequency $f$ as depicted in Fig.~\ref{expl}~(c). There are then several decay channels involving the formed dressed states (purple bars) from the upper to the lower manifold. In particular, the five transition frequencies $\omega_0$, $\omega_0 \pm (f + g)$ and $\omega_0 \pm (f - g)$ correspond to the five distinct frequencies of the Mollow quintuplet displayed in Fig.~\ref{specco}~(d). In the preceding qualitative discussion, the frequency $f$ is precisely the quantity appearing in the calculation of the dressed energy levels $\epsilon_{\pm, \pm}$ as described by Eq.~\eqref{eq:0sdsddssd1} and Eq.~\eqref{eq:0sdsddssd12323232}, and the five aforementioned transition frequencies arise from the differences between the various $\epsilon_{\pm, \pm}$.
\\


\section{\label{dissdiss}The regime of dissipative coupling}

In what follows, we focus on the purely dissipative ($\mathrm{ds}$) coupling regime, where the coherent coupling is set to zero ($g = 0$) as alluded to in the cartoon of Fig.~\ref{regimes}~(a). The average populations are considered in Sec.~\ref{disscoh}, the second-order degree of coherence in Sec.~\ref{disscorr}, and the power spectrum in Sec.~\ref{dissspec}. We especially focus on how distinctive features in this coupling regime arise as a function of the dissipative coupling strength $\gamma$ and driving strength $\Omega$ into the first 2LS.

This incoherent coupling regime can arise in several physical systems of current interest~\cite{Wang2020}, including in cavity optomechanical platforms~\cite{Qu2015}, with hybridized magnon-photon modes in cavities~\cite{Harder2018, Wang2019}, by exploiting photon Bose-Einstein condensates~\cite{Toebes2022}, and even with thermoacoustic oscillators~\cite{Doranehgard2022}.
\\


\subsection{\label{disscoh}Populations}

\begin{figure*}[tb]
 \includegraphics[width=\linewidth]{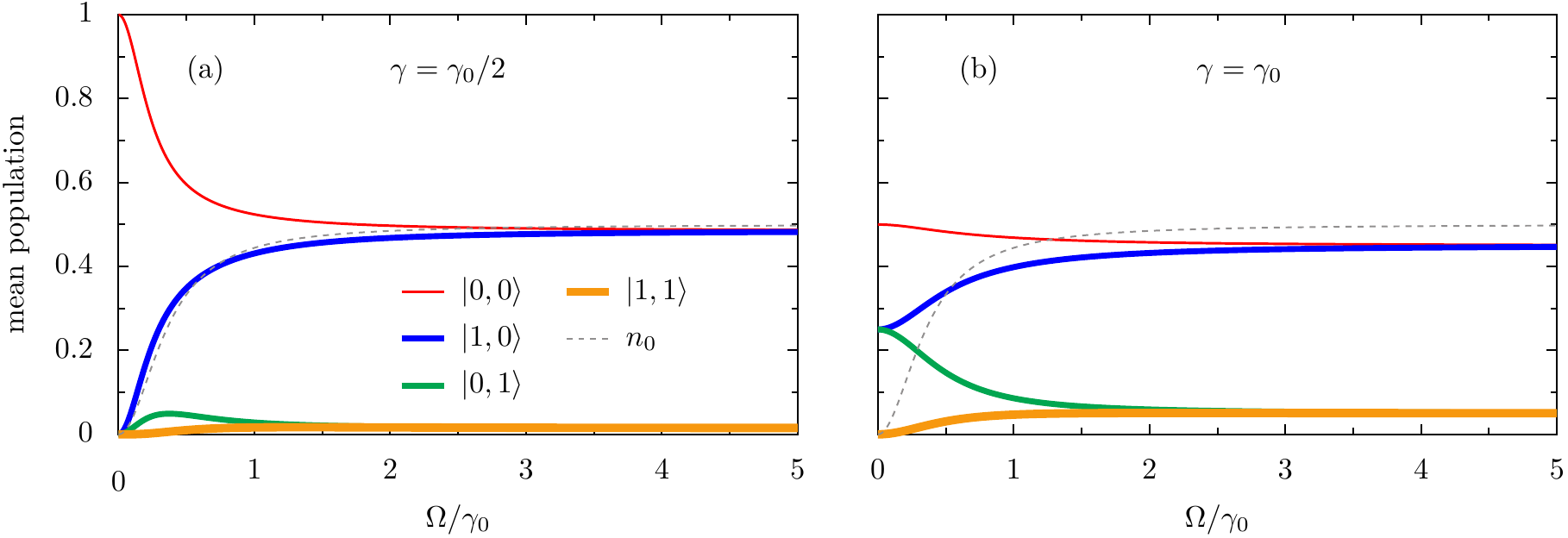}
 \caption{ \textbf{Populations in the dissipative coupling regime.} Average populations plotted against the drive amplitude $\Omega$, valid for when the incoherent coupling dominates [cf. Eq.~\eqref{eq:statespopgdiss}]. The single 2LS population is given by the dashed gray line [cf. Eq.~\eqref{eq:popuionsd343434issi}]. The incoherent coupling strength $\gamma = \gamma_0/2$ in Panel (a), while $\gamma = \gamma_0$ in Panel (b). The plot legend in panel (a) details the results for each state. }
 \label{popdiss}
\end{figure*}

The steady state mean populations $\rho_{n, m}^{\mathrm{ds}}$ of the four states $\ket{n, m}$ are (see Appendix~\ref{app:pop} for the theory) 
\begin{widetext}
\begin{subequations}
\label{eq:statespopgdiss}
\begin{align}
 \rho_{0, 0}^{\mathrm{ds}} &= \frac{ \left( 9 \gamma_0^2 - \gamma^2 \right) \left( \gamma_0^3 - \gamma_0 \gamma^2 \right)^2
+ 8 \gamma_0^2 \Omega^2 \left( 2 \gamma^4 - 3 \gamma^2 \gamma_0^2 + 9 \gamma_0^4 \right) 
+ 4 \Omega^4 \left( 44 \gamma_0^4 + 29 \gamma^2 \gamma_0^2 - 5 \gamma^4 \right)
+ 16 \Omega^6 \left( 8 \gamma_0^2 + \gamma^2 \right) }{ \left( 9 \gamma_0^2 - \gamma^2 \right) \left( \gamma_0^3 - \gamma_0 \gamma^2 \right)^2
+ 4 \gamma_0^2 \Omega^2 \left( 3 \gamma^4 + 2 \gamma^2 \gamma_0^2 + 27 \gamma_0^4 \right)
- 32 \Omega^4 \left( \gamma^2 - 10 \gamma_0^2 \right) \left( \gamma^2 + \gamma_0^2 \right)
+ 64 \Omega^6 \left( \gamma^2 + 4 \gamma_0^2 \right) }, \label{eq:vit1} \\
 \rho_{1, 0}^{\mathrm{ds}} &= \frac{ 4 \gamma_0 \Omega^2 \left( 9 \gamma_0^2 - \gamma^2 \right)
+ 4 \Omega^4 \left( 36 \gamma_0^4 + 25 \gamma^2 \gamma_0^2 - 3 \gamma^4 \right)
+ 16 \Omega^6 \left( \gamma^2 + 8 \gamma_0^2 \right) }{ \left( 9 \gamma_0^2 - \gamma^2 \right) \left( \gamma_0^3 - \gamma_0 \gamma^2 \right)^2
+ 4 \gamma_0^2 \Omega^2 \left( 3 \gamma^4 + 2 \gamma^2 \gamma_0^2 + 27 \gamma_0^4 \right)
- 32 \Omega^4 \left( \gamma^2 - 10 \gamma_0^2 \right) \left( \gamma^2 + \gamma_0^2 \right)
+ 64 \Omega^6 \left( \gamma^2 + 4 \gamma_0^2 \right) },  \label{eq:vit2} \\
 \rho_{0, 1}^{\mathrm{ds}} &= \frac{ 4 \gamma^2 \Omega^2 \left( 9 \gamma_0^4 + 9 \gamma_0^2 \Omega^2 + 4 \Omega^4 + \gamma^2 \left[ \Omega^2 - \gamma_0^2 \right] \right) }{ \left( 9 \gamma_0^2 - \gamma^2 \right) \left( \gamma_0^3 - \gamma_0 \gamma^2 \right)^2
+ 4 \gamma_0^2 \Omega^2 \left( 3 \gamma^4 + 2 \gamma^2 \gamma_0^2 + 27 \gamma_0^4 \right)
- 32 \Omega^4 \left( \gamma^2 - 10 \gamma_0^2 \right) \left( \gamma^2 + \gamma_0^2 \right)
+ 64 \Omega^6 \left( \gamma^2 + 4 \gamma_0^2 \right) }, \label{eq:vit3} \\
 \rho_{1, 1}^{\mathrm{ds}} &= \frac{ 4 \gamma^2 \Omega^4 \left( 4 \Omega^2 + 9 \gamma_0^2 - \gamma^2 \right) }{ \left( 9 \gamma_0^2 - \gamma^2 \right) \left( \gamma_0^3 - \gamma_0 \gamma^2 \right)^2
+ 4 \gamma_0^2 \Omega^2 \left( 3 \gamma^4 + 2 \gamma^2 \gamma_0^2 + 27 \gamma_0^4 \right)
- 32 \Omega^4 \left( \gamma^2 - 10 \gamma_0^2 \right) \left( \gamma^2 + \gamma_0^2 \right)
+ 64 \Omega^6 \left( \gamma^2 + 4 \gamma_0^2 \right) }. \label{eq:vit4}
\end{align} 
\end{subequations}
\end{widetext}
In the limiting case of strong driving ($\Omega \gg \gamma_0$), Eq.~\eqref{eq:statespopgdiss} collapses into the simples expressions
\begin{align}
\label{eq:statespopgenlimitdiss}
 \rho_{0, 0}^{\mathrm{ds}} &= \rho_{1, 0}^{\mathrm{ds}} = \frac{1}{2} \left( 1 - \frac{1}{2} \frac{(\gamma/2)^2}{(\gamma/2)^2 + \gamma_0^2} \right) ,
 \\ 
 \rho_{0, 1}^{\mathrm{ds}} &= \rho_{1, 1}^{\mathrm{ds}} = \frac{1}{4} \frac{(\gamma/2)^2}{(\gamma/2)^2 + \gamma_0^2},
\end{align}
which exhibit a coalescing of behaviors of the $\ket{0, 0}$ and $\ket{1, 0}$ states, and the $\ket{0, 1}$ and $\ket{1, 1}$ states respectively. Remarkably, in the opposing weak driving regime $\Omega \ll \gamma_0$, and with maximal dissipative coupling $\gamma \to \gamma_0$, we find from Eq.~\eqref{eq:statespopgdiss} the mean populations $\rho_{0, 0}^{\mathrm{ds}} = 1/2$, $\rho_{1, 0}^{\mathrm{ds}} = \rho_{0, 1}^{\mathrm{ds}} = 1/4$ and $\rho_{1, 1}^{\mathrm{ds}} = 0$. This analysis suggesting a population trapping effect, where the intermediate states $\ket{1, 0}$ and $\ket{0, 1}$ are associated with nonzero populations, which precipitates an unconventional ground state of the system. Notably, such a scenario is not possible for the case of purely coherently coupled 2LSs.

The average populations of Eq.~\eqref{eq:statespopgdiss} are plotted in Fig.~\ref{popdiss} as a function of the drive amplitude $\Omega$. Where the incoherent coupling strength $\gamma = \gamma_0/2$ in panel (a), the result is reminiscent of the coherently coupled case reported in Fig.~\ref{popco}~(a). Namely, the driven state population $\ket{1, 0}$ (medium blue line) is very similar to the single 2LS result (dashed gray line), while the populations of the undriven state $\ket{0, 1}$ (medium green line) and doubly-excited state (thick red line) are negligible. In Fig.~\ref{popdiss}~(b), where the dissipative coupling is maximal $\gamma = \gamma_0$, the surprising population trapping effect appears. Now, even in the limit of vanishing driving $\Omega \to 0$, the intermediate states of $\ket{0, 1}$ and $\ket{1, 0}$ (green and blue medium-thickness curves) are populated, such that the ground state (thin red line) is described by $\rho_{0, 0}^{\mathrm{ds}} = 1/2$ (and not $\rho_{0, 0}^{\mathrm{ds}} = 1$ as may have been reasonably envisaged). This population trapping effect has arisen due to the maximal dissipative coupling quenching transitions from the singly-excited states to the ground state. It has been noticed before within a dimer platform in an incoherent pumping setup~\cite{Gonzalez2011, MartinCano2011}.
\\


\subsection{\label{disscorr}Correlations}

\begin{figure}[tb]
 \includegraphics[width=1.0\linewidth]{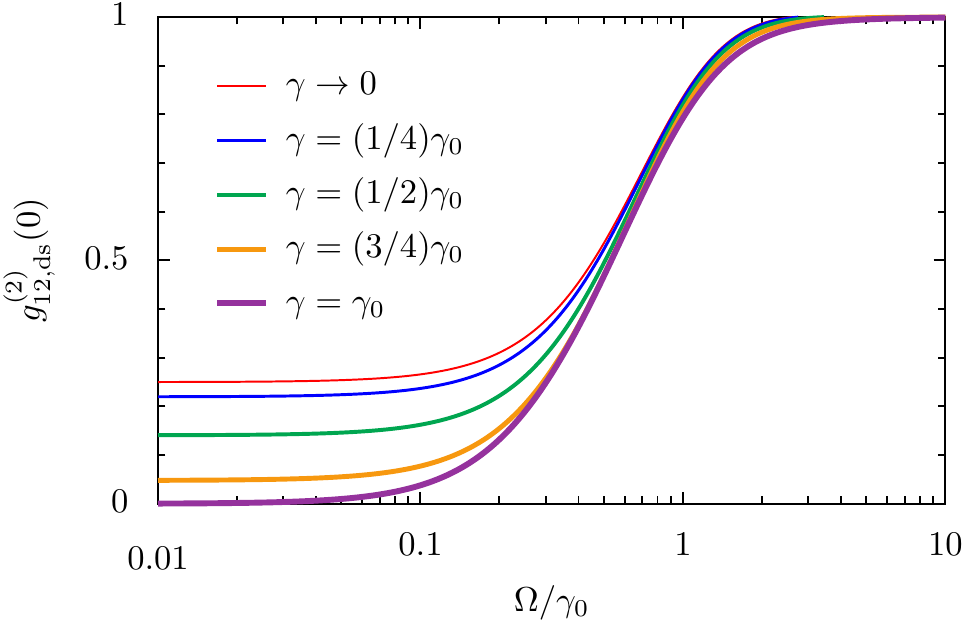}
 \caption{ \textbf{Correlations in the dissipative coupling regime.} Cross-correlator with zero time delay $g_{12, \mathrm{ds}}^{(2)} (0)$, plotted against the drive amplitude $\Omega$, valid for when the incoherent coupling dominates [cf. Eq.~\eqref{eq:g2diss}]. Stronger incoherernt coupling strengths $\gamma$ are associated with thicker lines. }
 \label{corrdiss}
\end{figure}

\begin{figure*}[tb]
 \includegraphics[width=\linewidth]{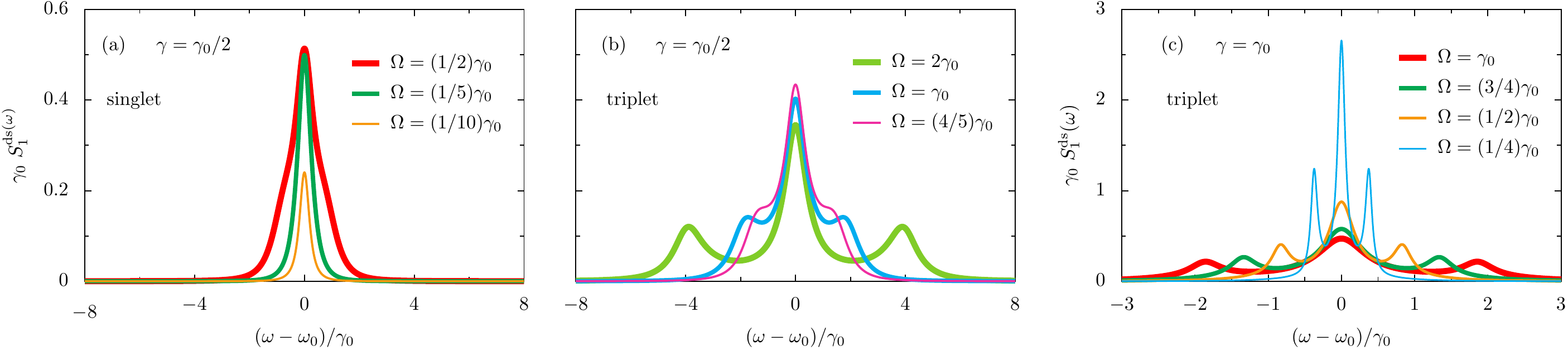}
 \caption{ \textbf{Spectra in the dissipative coupling regime.} Optical spectrum $S_1^{\mathrm{ds}}(\omega)$ of the first 2LS in the dissipative coupling regime, in units of the inverse damping rate $\gamma_0^{-1}$. In each panel, increasingly strong driving coupling $\Omega$ is denoted by increasingly thick colored lines. Panels (a) and (b): moderate dissipative coupling $\gamma = \gamma_0/2$, which displays singlet [panel~(a)] and triplet [panel~(b)] regimes. Panel (c): maximal dissipative coupling $\gamma = \gamma_0$ is associated with a spectral triplet only.  }
 \label{specdiss}
\end{figure*}

The expression for Glauber's (normalized) second-order cross-correlator at zero delay [cf. Eq.~\eqref{eq:0galub1}] in the incoherent coupling dominated arena reads (Appendix~\ref{app:pop} contains the underlying theory)
\begin{widetext}
\begin{equation}
\label{eq:g2diss}
 g_{12, \mathrm{ds}}^{(2)} (0) = 1 + \frac{ 1 }{ 4 } \frac{ \gamma^2 \left( \gamma^4 - 11 \gamma^2 \gamma_0^2 + 18 \gamma_0^4 - 2 \Omega^2 \left[ \gamma^2 + 6 \gamma_0^2 \right] \right)  }{ \gamma_0^4 \left( \gamma^2 - 9 \gamma_0^2 \right) + 2 \Omega^2 \left( 2 \gamma^4 - 17 \gamma^2 \gamma_0^2 - 18 \gamma_0^4 \right) - 8 \Omega^4 \left( \gamma^2 + 4 \gamma_0^2 \right) } + \frac{1}{4} \frac{ 4 \gamma_0^2 \Omega^2 - 27 \gamma_0^4 + \gamma^2 \left( 3 \gamma_0^2 - 10 \Omega^2 \right) }{ 9 \gamma_0^4 - \gamma^2 \gamma_0^2 + 18 \gamma_0^2 \Omega^2 + 8 \Omega^4 }.
\end{equation}
\end{widetext}
The asymptotics of Eq.~\eqref{eq:g2diss}, for both weak and strong driving, are as follows
\begin{align}
\label{eq:g2disslim}
 g_{12, \mathrm{ds}}^{(2)} (0) &= \frac{ \left( \gamma^2 - \gamma_0^2 \right)^2 }{ 4 \gamma_0^4 },
 \quad 
 &&\Omega \ll \gamma_0,  \\
  \quad 
 g_{12, \mathrm{ds}}^{(2)} (0) &= 1,
 \quad 
&& \Omega \gg \gamma_0. 
\end{align}
Therefore, with weak coupling the dissipative coupling regime presents anticorrelations $g_{12, \mathrm{ds}}^{(2)} (0) < 1$, and minimally $g_{12, \mathrm{ds}}^{(2)} (0) \to 0$ when the incoherent coupling is maximal $\gamma \to \gamma_0$. Strong driving $\Omega \gg \gamma_0$ wipes out any correlations since the system behaves as if it is uncoupled.

We plot $g_{12, \mathrm{ds}}^{(2)} (0)$ in Fig.~\ref{corrdiss} as a function of the drive amplitude $\Omega$. Stronger incoherent couplings $\gamma$ are marked with thicker lines. As for the coherently coupled case of Fig.~\ref{corrco}, strong driving $\Omega \gg \gamma_0$ washes out any correlations $g_{12, \mathrm{ds}}^{(2)} (0) \to 1$, independent of the dissipative coupling strength $\gamma$. Most noticeably, between $\Omega \simeq \gamma_0$ and $\Omega \simeq \gamma_0/10$, the cross-correlator reaches its minimum, which is highly dependent on the dissipative coupling strength $\gamma$. For the case of very small incoherent coupling $\gamma \to 0$ (thinnest, red line), the minimum of $g_{12, \mathrm{ds}}^{(2)} (0) \to 1/4$ is reached, while for maximal dissipative coupling $\gamma \to \gamma_0$ (thick purple line), the absolute minimum of $g_{12, \mathrm{ds}}^{(2)} (0) \to 0$ is arrived at, making the dissipatively coupled regime highly discriminatory within these statistics.


\subsection{\label{dissspec}Spectrum}

In Fig.~\ref{specdiss}, we plot the optical spectrum of the first 2LS in the dissipative coupling regime $S_1^{\mathrm{ds}}(\omega)$ [cf. Eq.~\eqref{eq:speccccc}]. In each panel, increasingly strong driving coupling $\Omega$ is denoted by increasingly thick colored lines, as indicated by the legends. Importantly, in Fig.~\ref{specdiss} panels (a) and (b) the dissipative coupling strength is below its maximum value, being $\gamma = \gamma_0/2$. With weak driving $\Omega \le \gamma_0/2$ in panel (a), we see a simple singlet spectral structure, with hints of other, unresolvable peaks for higher driving strengths (thicker lines). In panel (b), with strong driving $\Omega > \gamma_0/2$, a distinctive triplet structure has finally emerged, which is particularly visible for $\Omega > \gamma_0$ (blue and green thicker line). In this way, the dissipatively coupled regime presents a much simpler behavior than the coherently coupled regime, where quintuplets may also be formed. This is because the absence of any coherent coupling ($g = 0$) ensures that all spectral peaks arise from the driving only, in a similar way to the case of a single 2LS (see Appendix~\ref{app:single_2LS} for the interpretation of the spectrum of one 2LS) so that only spectral singlets or spectral triplets may form, depending upon whether the driving is above or below a threshold dissipation rate (which here is composed of contributions from both $\gamma$ and $\gamma_0$).

In Fig.~\ref{specdiss}~(c), where the dissipative strength is increased to its maximal strength of $\gamma = \gamma_0$, a very different spectral evolution is displayed. Now, a triplet structure is observed for all driving strengths, since the weightings of key transitions [those which allowed for the spectral singlet-triplet transition in panels (a) and (b)] are proportional to $\gamma_0 - \gamma$ and thus disappear in this parameter regime [in a similar way, the population trapping effect reported in Fig.~\ref{popdiss}~(b) only occurs when $\gamma \to \gamma_0$]. Of course, the triplet displayed in Fig.~\ref{specdiss}~(c) becomes more widely separated with increasing driving strengths (thicker lines), and the persistence of the spectral triplet acts as an indicator of maximal dissipative coupling being reached.
\\


\section{\label{uniuni}The regime of unidirectional coupling}

\begin{figure*}[tb]
 \includegraphics[width=\linewidth]{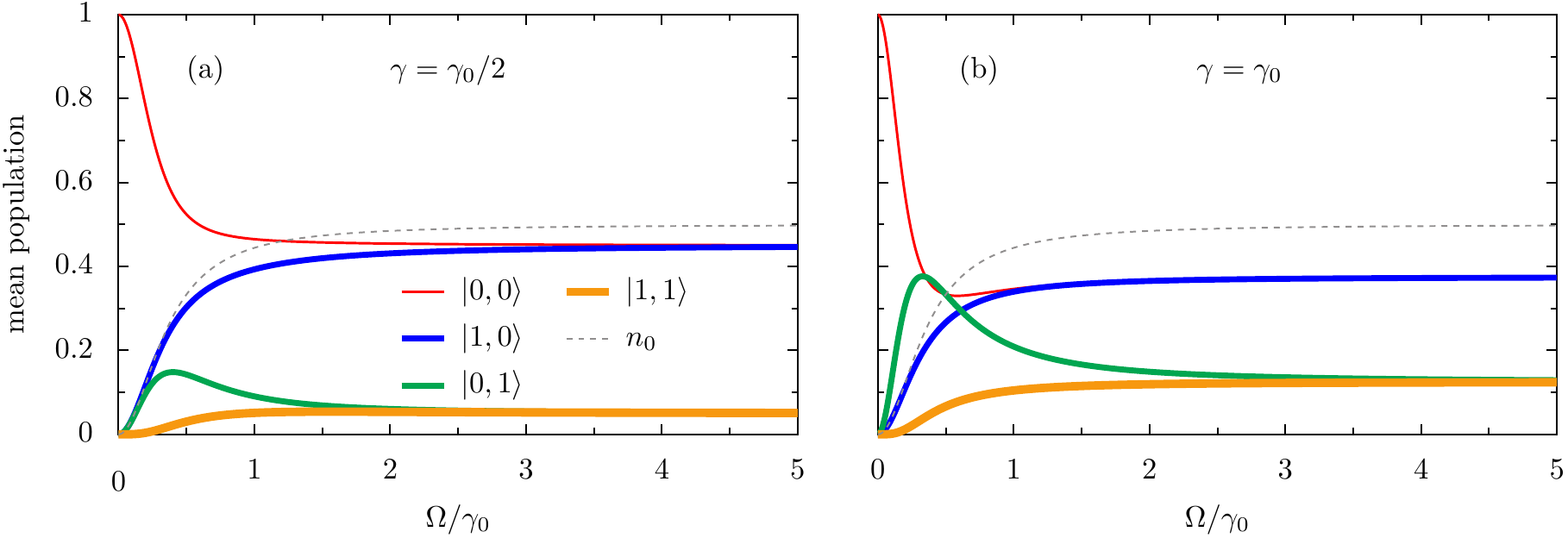}
 \caption{ \textbf{Populations in the unidirectional coupling regime.} Mean populations in the unidirectional coupling regime, as a function of the drive amplitude $\Omega$, in units of the damping rate $\gamma_0$ [Eq.~\eqref{eq:statespopguni}]. Dashed gray line: single 2LS population [cf. Eq.~\eqref{eq:popuionsd343434issi}]. Panel (a): moderate dissipative coupling strength $\gamma = \gamma_0/2$. Panel (b): maximal dissipative coupling strength $\gamma = \gamma_0$. The labeling of the mean population of the state $\ket{i, j}$ is displayed in the legend of panel (a), and states with a number of excitations $N=\{ 0, 1, 2\}$ are shown with increasingly thick lines.}
 \label{popuni}
\end{figure*}

Here we consider the unidirectional coupling regime, where the coherent and dissipative coupling strengths are fixed by the ratio $g/\gamma = 1/2$, as enforced by Eq.~\eqref{eq:conditions2}. The relative phase $\theta-\phi$ can be equal to $\pi/2$ or $3\pi/2$, as follows from Eq.~\eqref{eq:conditions1}, depending if the one-way coupling between the pair of 2LSs follows the directionality $1 \rightarrow 2$ or $1 \leftarrow 2$, as sketched in Fig.~\ref{regimes}~(c). Within this one-way coupling regime, we discuss the mean populations in Sec.~\ref{unicoh}, the second-order degree of coherence in Sec.~\ref{unicorr}, and the optical spectrum in Sec.~\ref{unispec}. We concentrate on the case of unidirectionality in the direction of $1 \rightarrow 2$ only in this section since only the first 2LS is being driven.

This unusual coupling regime can be realized in a number of platforms, including with a metallic nanoparticle and a quantum emitter housed inside a cavity~\cite{Tan2022}, with metal-insulator-metal heterostructures~\cite{Li2022}, using a couple of yttrium iron garnet microspheres~\cite{Zhan2022}, and by exploiting a pair of quantum emitters in a suitable dielectric environment~\cite{Miguel2022}.

\subsection{\label{unicoh}Average populations}

The average populations $\rho_{n, m}^{\mathrm{un}}$ associated with the four states $\ket{n, m}$ in the unidirectional ($\mathrm{un}$) regime $1 \rightarrow 2$ [cf. Fig.~\ref{regimes}~(c)] read (see Appendix~\ref{app:pop} for the theoretical background)
\begin{widetext}
\begin{subequations}
\label{eq:statespopguni}
\begin{align}
 \rho_{0, 0}^{\mathrm{un}} &= \frac{1}{\gamma_0^2 + 8 \Omega^2} \frac{ 9 \gamma_0^8 + 8 \gamma_0^4 \Omega^2 \left( 9 \gamma_0^2 - 4 \gamma^2 \right) + 16 \Omega^4 \left( 11 \gamma_0^4 + 21 \gamma^2 \gamma_0^2 - 4 \gamma^4 \right) + 64 \Omega^6 \left( 2 \gamma_0^2 + \gamma^2 \right) }{ 9 \gamma_0^6 + 4 \gamma_0^2 \Omega^2 \left( 28 \gamma^2 + 9 \gamma_0^2 \right) + 32 \Omega^4 \left( \gamma^2 + \gamma_0^2 \right) }, \label{eq:vitsd1} \\
 \rho_{1, 0}^{\mathrm{un}} &= \frac{4 \Omega^2}{\gamma_0^2 + 8 \Omega^2} \left( 1 - \frac{ 4 \gamma^2 \Omega^2 \left( 9 \gamma_0^2 + 4 \Omega^2 \right) }{ 9 \gamma_0^6 + 4 \gamma_0^2 \Omega^2 \left( 28 \gamma^2 + 9 \gamma_0^2 \right) + 32 \Omega^4 \left( \gamma^2 + \gamma_0^2 \right) } \right),  \label{eq:visdt2} \\
 \rho_{0, 1}^{\mathrm{un}} &= \frac{ 16 \gamma^2 \Omega^2 }{ \gamma_0^2 + 8 \Omega^2 } \frac{ 9 \gamma_0^4 + 9 \gamma_0^2 \Omega^2 + 4 \Omega^2 \left( \gamma^2 + \Omega^2 \right) }{ 9 \gamma_0^6 + 4 \gamma_0^2 \Omega^2 \left( 28 \gamma^2 + 9 \gamma_0^2 \right) + 32 \Omega^4 \left( \gamma^2 + \gamma_0^2 \right) }, \label{eq:visdt3} \\
 \rho_{1, 1}^{\mathrm{un}} &= \frac{ 16 \gamma^2 \Omega^4 }{ \gamma_0^2 + 8 \Omega^2 } \frac{ 9 \gamma_0^2 + 4 \Omega^2 }{ 9 \gamma_0^6 + 4 \gamma_0^2 \Omega^2 \left( 28 \gamma^2 + 9 \gamma_0^2 \right) + 32 \Omega^4 \left( \gamma^2 + \gamma_0^2 \right) }. \label{eq:visdsdt4}
\end{align} 
\end{subequations}
\end{widetext}
The hallmark of unidirectional coupling is seen through the exact relation $\rho_{1, 0}^{\mathrm{un}} + \rho_{1, 1}^{\mathrm{un}} = n_0$. That is, the mean population of 2LS-1 is exactly that of a solitary, uncoupled 2LS [cf. Eq.~\eqref{eq:popuionsd343434issi}] due to the backaction from 2LS-2 being nullified by the mixture of coherent and incoherent coupling. In the limit of strong driving ($\Omega \gg \gamma_0$), Eq.~\eqref{eq:statespopguni} reduces to
\begin{align}
\label{eq:statespopgenlimitduni}
 \rho_{0, 0}^{\mathrm{un}} = \rho_{1, 0}^{\mathrm{un}} &= \frac{1}{4} \frac{ \gamma^2 +  2 \gamma_0^2 }{\gamma^2 + \gamma_0^2}, \nonumber \\
 \rho_{0, 1}^{\mathrm{un}} = \rho_{1, 1}^{\mathrm{un}} &= \frac{1}{4} \frac{ \gamma^2 }{\gamma^2 + \gamma_0^2},
\end{align}
such that the four state populations again couple into pairs, in the same manner as for the coherent and dissipative coupling regimes of the preceding sections.

\begin{figure*}[tb]
 \includegraphics[width=1.0\linewidth]{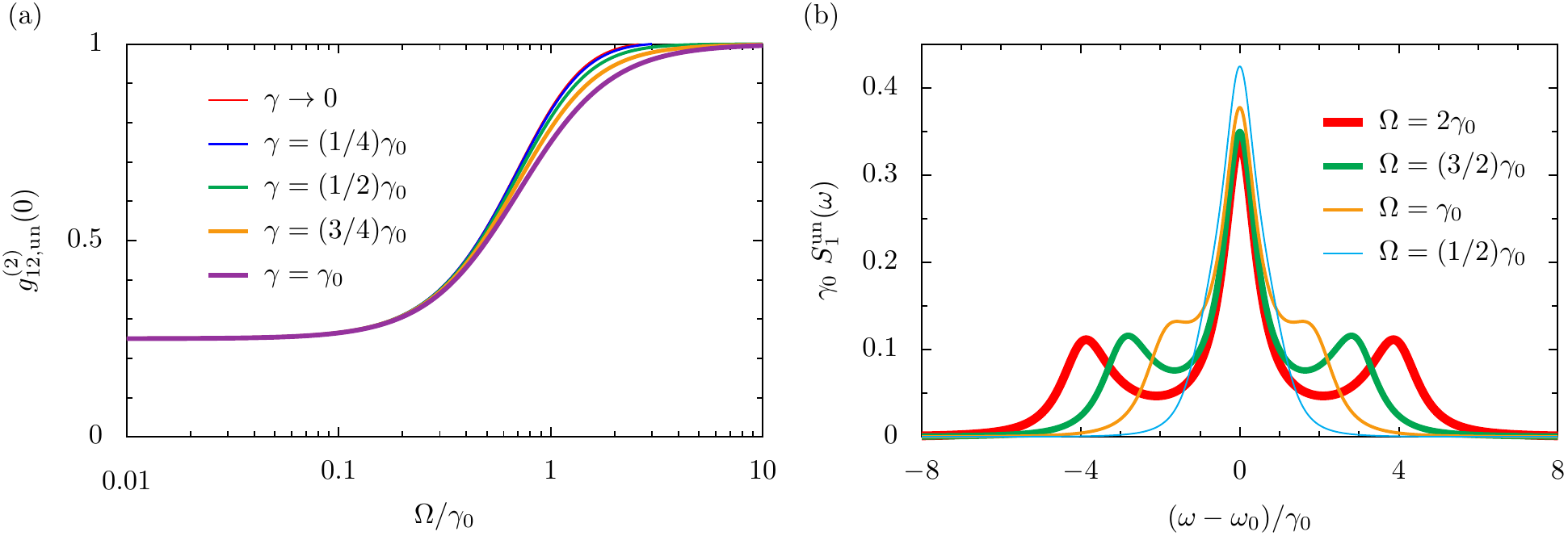}
 \caption{ \textbf{Correlations and spectra in the unidirectional coupling regime.} Panel (a): Cross-correlator for the case of zero time delay $g_{12, \mathrm{un}}^{(2)} (0)$, plotted against the drive amplitude $\Omega$ and when unidirectional effective interactions dominate. Stronger incoherent couplings $\gamma$ are associated with thicker curves. Panel (b): Power spectrum $S_1^{\mathrm{un}}(\omega)$ of the first 2LS in the unidirectional coupling regime. Stronger driving coupling $\Omega$ is denoted by thicker coloured curves. The incoherent coupling $\gamma = \gamma_0$ and the unidirectional conditions of Eq.~\eqref{eq:conditions} are observed. }
 \label{specuni}
\end{figure*}

We plot in Fig.~\ref{popuni} the average populations during the unidirectional coupling dominance, against the drive amplitude $\Omega$. The dashed gray line is the single 2LS population [cf. Eq.~\eqref{eq:popuionsd343434issi}]. In panel (a), where the incoherent coupling $\gamma = \gamma_0/2$, the evolution of the average populations is similar to the dissipatively coupled regime of Fig.~\ref{popdiss}~(a), but with a more noticeable rise in population of the state $\ket{0, 1}$ (medium green line) because of the unidirectional character of the effective interaction from two-level system $1 \rightarrow 2$. In Fig.~\ref{popuni}~(b), where the dissipative coupling is maximal ($\gamma = \gamma_0$), the peak in population of the state $\ket{0, 1}$ (medium green line) is even more apparent due to the strongest possible unidirectional coupling in its favour, such that it is even the most populated state for $\Omega \lesssim \gamma_0/2$. For stronger driving $\Omega > (3/4) \gamma_0$, the unidirectional coupling is finally overcome by the coherent driving into 2LS-1, and the $\ket{1, 0}$ (medium blue line) becomes the most likely state to be populated.


\subsection{\label{unicorr}Correlations}

The expression for Glauber's (normalized) second-order cross-correlator at zero delay $g_{12, \mathrm{un}}^{(2)} (0)$ in the unidirectional coupling regime [cf. Eq.~\eqref{eq:0galub1}] reads (see Appendix~\ref{app:pop} for the supporting theory)
\begin{equation}
\label{eq:g2uniii}
 g_{12, \mathrm{un}}^{(2)} (0) = \frac{ \left( \gamma_0^2 + 8 \Omega^2 \right) \left( 9 \gamma_0^2 + 4 \Omega^2 \right) }{ 36 \gamma_0^2 + 8 \Omega^2 \left( 2 \gamma^2 + 9 \gamma_0^2 \right) + 32 \Omega^4 },
\end{equation} 
which has the following simple limiting forms in the small and large driving limits
\begin{align}
\label{eq:g2unizscii}
 g_{12, \mathrm{un}}^{(2)} (0) &= \tfrac{ 1 }{ 4  },
 \quad
 &&\Omega \ll \gamma_0, 
 \\
 g_{12, \mathrm{coI}}^{(2)} (0) &= 1,
 \quad  
&& \Omega \gg \gamma_0. 
\end{align}
The cross-correlator $g_{12, \mathrm{un}}^{(2)} (0)$ is displayed in Fig.~\ref{specuni}~(a), and is plotted against the drive amplitude $\Omega$. Stronger incoherent couplings $\gamma$ are associated with thicker curves. Most noticeably, the figure shows universal anticorrelations $g_{12, \mathrm{un}}^{(2)} (0) < 1$. The variation in dissipative coupling strength $\gamma$ is not very discriminatory in this quantity, as it sweeps between the asymptotic values of $1/4$ and $1$ in a sigmoid-function-like manner.

\subsection{\label{unispec}Spectrum}

In Fig.~\ref{specuni}~(b), we display the optical spectrum [cf. Eq.~\eqref{eq:speccccc}] of the first 2LS in the unidirectional coupling regime $S_1^{\mathrm{un}}(\omega)$. As expected, it indeed displays all of the hallmarks of the spectrum of a single 2LS because of the absence of any backaction from the second 2LS~\cite{Camilo2016b}. It thus follows the exact expression familiar from the single 2LS case [cf. Appendix~\ref{app:single_2LS} for the derivation]
\begin{widetext}
\begin{equation}
\label{eq:nmycdfsoef2sdsd2323}
S_1^{\mathrm{un}}(\omega) =  \frac{1}{2 \pi} \frac{ \tfrac{\gamma_0}{2} }{ ( \tfrac{\gamma_0}{2} )^2 + \left( \omega - \omega_0 \right)^2 }  - \frac{\gamma_0}{\pi} \frac{ \left( \omega - \omega_0 \right)^2 + \gamma_0^2 - 16 \Omega^2}{4 \left( \omega - \omega_0 \right)^4 + \left( \omega - \omega_0 \right)^2 \left( 5 \gamma_0^2 - 32 \Omega^2 \right) + \left( \gamma_0^2 + 8 \Omega^2 \right)^2 }.
\end{equation}
\end{widetext}
Notably, as show in Fig.~\ref{specuni}~(b), there is a singlet regime with weak driving (thin cyan line), before a Mollow triplet regime arises with strong driving (thicker yellow, green and red lines), in a remarkable manifestation of single atom physics within an ostensibly coupled pair system. Due to the one-way nature of the coupling, the interpretation of this triplet structure is essentially the same as for the one atom case (see Fig.~\ref{tran} and the discussion around it for example). The unidirectional coupling regime discussed in this section is the most extreme example of a more general asymmetric coupling, and for completeness we discuss a few results with less extreme asymmetric coupling in Appendix~\ref{gengen}.
\\


\section{\label{conc}Conclusions}

We have completed a systematic survey of some of the fundamental quantum optical properties of an asymmetrically driven-dissipative pair of two-level systems. In an open quantum systems approach, we have calculated the mean populations, coherences, and optical spectra which this simple setup supports. We have revealed how these key quantities change depending on the location within the coupling landscape, which encompasses coherent, dissipative, unidirectional, and asymmetric coupling. In particular, we have shown how the celebrated Mollow triplet spectrum for a single atom can be reproduced in a pair system with unidirectional coupling, and how away from this special regime anything from a singlet to a quintuplet may appear. We have also reported an instance where the population is trapped (in the limit of maximal incoherent interactions) which is associated with the disappearance of the spectral singlet to triplet transition, and we have revealed the gamut of strong correlations appearing in the duo of atoms. 

Our basic theory, spanning both coherent and incoherent coupling, may be realized in a plethora of artificial atom systems~\cite{Chang2018}, including with ultracold atoms~\cite{Cooper2019}, superconducting qubits~\cite{Kjaergaard2020}, and plasmonic nanoparticles~\cite{Bordo2019,Downing2017}. The presented theory opens up the opportunity for the detection of cooperative Mollow physics and chiral physics within an elemental dimer system, with natural extensions. Indeed, the scaling up of the dimer system into a quantum network may present novel opportunities for quantum transport and communication, as well as for quantum information processing.  
\\
\\


\noindent
\textbf{Acknowledgements}\\
\textit{Funding:} CAD is supported by a University Research Fellowship (URF\slash R1\slash 201158) from the Royal Society. EdV acknowledges the CAM Pricit Plan (Ayudas de Excelencia del Profesorado Universitario), the TUMIAS Hans Fischer Fellowship, and the MCIN/AEI via Grant No. PID2020-113415RB-C22. AIFD acknowledges sponsorship from the Spanish MCIN/AEI via the Grant Nos. PID2021-126964OB-I00 and TED2021-130552B-C22, as well as the European Union's Horizon Europe Research and Innovation Programme under agreement 101070700. \textit{Discussions:} We are grateful to J.~C.~L\'{o}pez~Carre\~{n}o for discussions in the early stages of this work.\\\\
\\


\appendix

\renewcommand{\theequation}{A \arabic{equation}}
\section{A single driven-dissipative two-level system}
\label{app:single_2LS}


Here we detail some explanatory results for a single two-level system (2LS) driven coherently~\cite{Mollow1969, Breuer2002, delVallePRL2010,ValleLaussy2011, Camilo2016b}. We describe the open quantum systems model in Sec.~\ref{app:single_2LS_model}, before studying the populations and coherences in Sec.~\ref{app:one_time} and the optical spectrum in Sec.~\ref{app:spec}. Notably, the few-level system considered here ensures that the Hilbert space is very small, and that the equations of motion naturally close with the second moments.
\\


\subsection{Model}
\label{app:single_2LS_model}

\begin{figure}[tb]
 \includegraphics[width=1.0\linewidth]{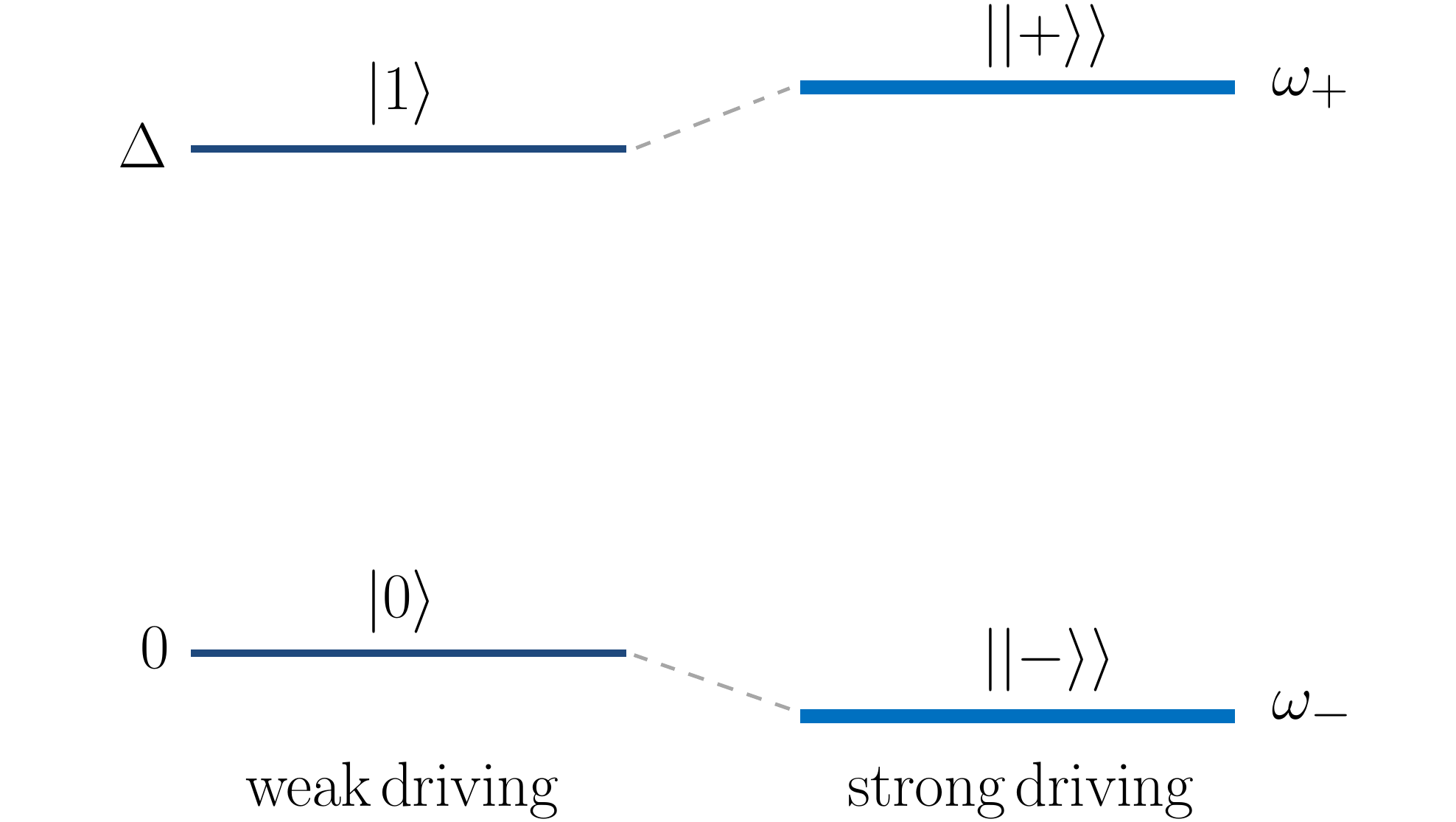}
 \caption{ \textbf{A sketch of the energy ladder of a single 2LS.} A 2LS of resonance frequency $\omega_0$, driven coherently with the driving amplitude $\Omega$ and frequency $\omega_{\mathrm{d}}$ (the detuning $\Delta = \omega_0 - \omega_{\mathrm{d}}$). We show both the weak (left) and strong (right) driving regimes [cf. Eq.~\eqref{eq:appjsdsdhjsdsd2}].}
 \label{singlesketch}
\end{figure}

The Hamiltonian $H$ of a single, coherently driven 2LS is composed of two contributions as follows
\begin{equation}
\label{eq:app00}
 H = H_0 + H_{\mathrm{d}},
\end{equation}
where the 2LS excitation term $H_0$ and the drive term $H_{\mathrm{d}}$ are given by
\begin{align}
\label{eq:app01}
 H_0 &= \omega_0 \sigma^{\dagger} \sigma, 
\\
 H_{\mathrm{d}} &= \Omega \left( \mathrm{e}^{\mathrm{i} \omega_{\mathrm{d}} t} \sigma + \mathrm{e}^{- \mathrm{i} \omega_{\mathrm{d}} t} \sigma^{\dagger} \right),
\end{align}
where the 2LS is characterized by the transition frequency $\omega_0$, and the coherent drive has the amplitude $\Omega$ and the frequency $\omega_{\mathrm{d}}$. The 2LS raising and lowering operators $\sigma^{\dagger}$ and $\sigma$ act such that $\sigma^{\dagger} \ket{0} = \ket{1}$ and $\sigma \ket{1} = \ket{0}$. The 2LS has a two-dimensional Hilbert space, composed of the states $\ket{0}$ and $\ket{1}$, and upon neglecting the drive term $H_{\mathrm{d}}$ the associated eigenfrequencies are simply $0$ and $\omega_0$. Upon moving to a rotating frame, the Hamiltonian of Eq.~\eqref{eq:app00} transforms into $\tilde{H} = U H U^{\dagger} + \mathrm{i} \left( \partial_{t} U \right) U^{\dagger}$. With the operator $U = \mathrm{e}^{\mathrm{i} \omega_{\mathrm{d}} t \sigma^{\dagger} \sigma}$, one finds the transformed Hamiltonian $ \tilde{H}$ becomes
\begin{equation}
\label{eq:app02}
 \tilde{H} = \Delta \sigma^{\dagger} \sigma + \Omega \left( \sigma + \sigma^{\dagger} \right),
\end{equation}
with the detuned frequency $\Delta = \omega_0 - \omega_{\mathrm{d}}$. In the bare state basis $\{ \ket{0}, \ket{1} \}$ , the $2 \times 2$ matrix representation of the transformed Hamiltonian $ \tilde{H}$ reads
\begin{equation}
\label{eqapp:Ham_rep}
\tilde{H} =
\begin{pmatrix}
  0 &  \Omega \\
  \Omega & \Delta
 \end{pmatrix},
 \end{equation}
which suggests the following diagonalized form of the Hamiltonian
\begin{equation}
\label{eq:appjhjsdsd2}
 \tilde{H} = \omega_+ \kettwo{+} \bratwo{+} + \omega_- \kettwo{-} \bratwo{-}.
\end{equation}
Here the two dressed energy levels $\omega_{\pm}$ read
\begin{equation}
\label{eq:appjsdsdhjsdsd2}
 \omega_{\pm} = \frac{\Delta}{2} \pm R,
\end{equation}
where we have introduced the splitting frequency $R$, defined via the expression
\begin{equation}
\label{eq:RRRRRRRRappjsdsdhjsdsd2}
  R = \sqrt{ \left( \frac{\Delta}{2} \right)^2 + \Omega^2 }.
\end{equation}
The two dressed eigenstates $\kettwo{\pm}$ appearing in Eq.~\eqref{eq:appjhjsdsd2}, and corresponding to the eigenfrequencies $\omega_{\pm}$, read
\begin{align}
\label{eq:appjsddfdsdhjsdasdsd2}
 \kettwo{+} &= \sin \theta \ket{0} + \cos \theta \ket{1}, \\
  \kettwo{-} &= \cos \theta \ket{0} - \sin \theta \ket{1},
\end{align}
where we use single (double) kets for the bare (dressed) eigenstates. In Eq.~\eqref{eq:appjsddfdsdhjsdasdsd2}, the two Bogoliubov coefficients are defined by
\begin{align}
\label{eq:appjsddsdsdsfdsdhjsdsd2}
 \sin \theta &= \frac{1}{\sqrt{2}} \sqrt{ 1 - \frac{\Delta}{2 R} }, \\
 \cos \theta &= \frac{1}{\sqrt{2}} \sqrt{ 1 + \frac{\Delta}{2 R} }.
\end{align}
In the limit of weak driving ($\Omega \to 0$, and so $R \to \Delta/2$), one recovers the undressed results since $\kettwo{+} \to \ket{1}$ and $\kettwo{-} \to  \ket{0}$, while the eigenfrequencies satisfy $\omega_+ \to \Delta$ and $\omega_- \to 0$.

We plot the energy ladder of the 2LS in Fig.~\ref{singlesketch} in both the weak (left) and strong (right) driving regimes [cf. Eq.~\eqref{eq:appjsdsdhjsdsd2}]. It illustrates the enlarged inter-level frequency splitting from $\Delta$ to $\omega_+ - \omega_- = 2 R$ due to the coherent driving of strength $\Omega$, which is important for the discussion of the optical spectrum later on.

The quantum master equation of the system's density matrix $\rho$ is considered to be in the typical Lindblad form~\cite{Gardiner2014}
\begin{equation}
\label{eqapp:master}
 \partial_t \rho = \mathrm{i} [ \rho, \tilde{H} ] +  \frac{\gamma}{2} \left( 2 \sigma \rho \sigma^{\dagger} -  \sigma^{\dagger} \sigma \rho - \rho \sigma^{\dagger} \sigma \right),
\end{equation}
in terms of the transformed Hamiltonian $\tilde{H}$ of Eq.~\eqref{eq:app02}, and where $\gamma$ is the damping decay rate of the 2LS. Equation~\eqref{eqapp:master}, in conjunction with the quantum regression formula, gives rise to the results of the following subsections for the mean population and coherence [Sec.~\ref{app:one_time}] and optical spectrum [Sec.~\ref{app:spec}].
\\


\subsection{Population and coherence}
\label{app:one_time}

\begin{figure*}[tb]
 \includegraphics[width=\linewidth]{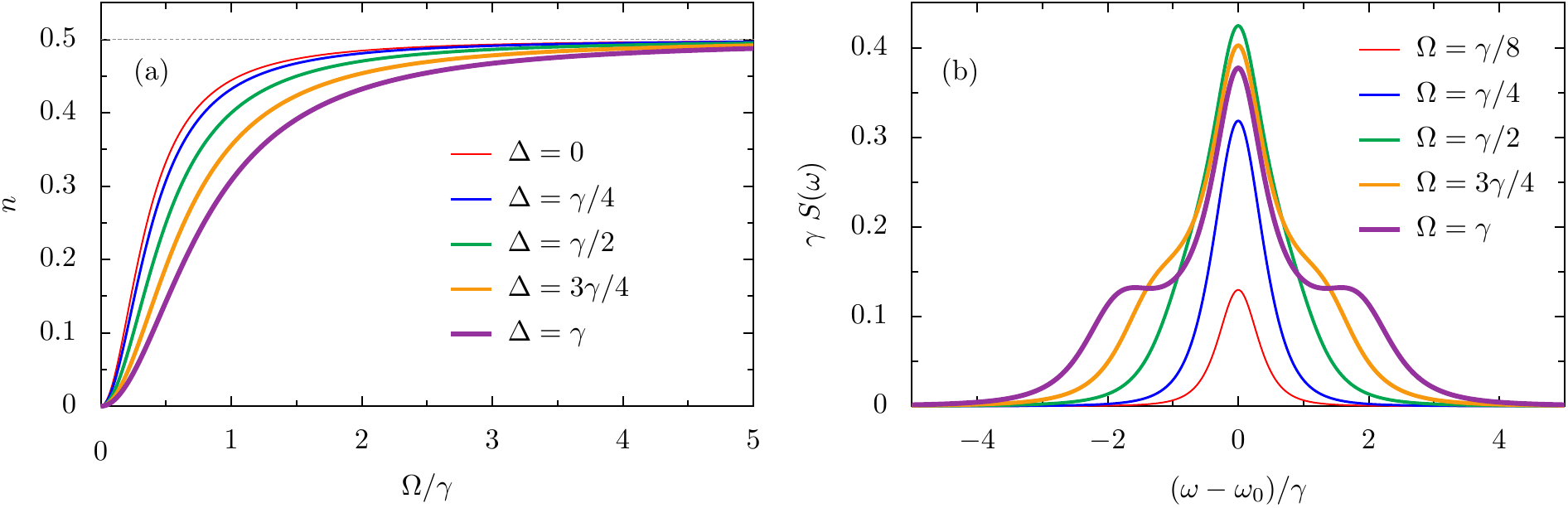}
 \caption{ \textbf{Population and spectrum of a single 2LS.} Panel (a): The mean steady state population $n$ of a single 2LS as a function of the drive strength $\Omega$, in units of the damping rate $\gamma$ [cf. Eq.~\eqref{eq:populationsd343434issi}]. We show results for increasingly strong detuning frequencies $\Delta$ (increasingly thick lines). Dashed gray line: guide for the eye at $1/2$. Panel (b): The spectrum of a single 2LS, as a function of the frequency $\omega$, in units of the damping rate $\gamma$ [cf. Eq.~\eqref{eq:nmycoef2sdsd2323}]. We consider zero detuning [$\Delta = 0$, corresponding to the thin red line in panel (a)] and show results for increasingly large drive amplitudes $\Omega$ (increasingly thick lines).}
 \label{single}
\end{figure*}

The quantum regression formula and the quantum master equation of Eq.~\eqref{eqapp:master} together yield the following equation of motion for the mean values of the first and second moments of the 2LS,
\begin{equation}
\label{eqapp:of_motion}
\partial_t \mathbf{u}  = \mathbf{P} - \mathbf{Q} \mathbf{u},
\end{equation}
for the 3-vector of correlators $\mathbf{u}$, and with the drive term $\mathbf{P}$ and the dynamical matrix $\mathbf{Q}$, where these objects are defined by
\begin{equation}
\label{eqapp:umatrix}
\mathbf{u}  =
\begin{pmatrix}
  \langle \sigma  \rangle   \\
  \langle {\sigma^{\dagger}} \rangle  \\
  \langle {\sigma^{\dagger}} \sigma \rangle 
 \end{pmatrix},
 \quad\quad\quad 
\mathbf{P} = \mathrm{i} \Omega
\begin{pmatrix}
  -1  \\
  1 \\
    0 
 \end{pmatrix},
 \end{equation}
 \begin{equation}
\label{eqapp:umatrix543453t} 
 \mathbf{Q} =
\begin{pmatrix}
  \tfrac{\gamma}{2} + \mathrm{i} \Delta &  0 &  - 2 \mathrm{i} \Omega \\
   0 & \tfrac{\gamma}{2} - \mathrm{i} \Delta  & 2\mathrm{i} \Omega \\
   -\mathrm{i} \Omega & \mathrm{i} \Omega & \gamma
 \end{pmatrix}.
 \end{equation}
In the steady state ($\mathrm{ss}$), where at long times ($t \to \infty$) the first time derivative $\partial_t \mathbf{u} = 0$, it follows from Eq.~\eqref{eqapp:of_motion} that the steady state population of a single 2LS, $n = \langle {\sigma^{\dagger}} \sigma \rangle_{\mathrm{ss}}$, and its coherence $c = \langle \sigma \rangle_{\mathrm{ss}}$, are captured by the expressions
\begin{align}
\label{eq:populationsd343434issi}
 n &= \frac{ \Omega^2}{2 \Omega^2 + \Delta^2 + \left( \tfrac{\gamma}{2} \right)^2 }, 
\\
 c &= -\frac{ \Omega \left( \Delta + \mathrm{i} \tfrac{\gamma}{2} \right) }{2 \Omega^2 + \Delta^2 + \left( \tfrac{\gamma}{2} \right)^2 }.  
\end{align}
For the case of zero detuning ($\Delta = 0$), Eq.~\eqref{eq:populationsd343434issi} reproduces the mean population given as $n_0$ in Eq.~\eqref{eq:popuionsd343434issi} in the main text (where we also relabel $\gamma$ as $\gamma_0$ to match the notation of the dimer problem). Notably, in the limit of strong driving $\Omega \gg \gamma$, the mean population reaches its maximum value of $n = 1/2$, since population inversion cannot occur with purely coherent driving~\cite{Camilo2016b}.

We plot the steady state population $n$ in Fig.~\ref{single}~(a), as a function of the drive strength $\Omega$, for increasing large detuning frequencies $\Delta$ (increasingly thick colored lines). The plot displays graphically the evolution of $n$ up to the threshold value of $n = 1/2$, revealing the saturation in the system for large drivings.
\\


\subsection{Spectrum}
\label{app:spec}

Applying the quantum regression formula along with the quantum master equation of Eq.~\eqref{eqapp:master} leads to a two-time equation of motion [cf. Eq.~\eqref{eqapp:of_motion}] containing the object $\langle {\sigma^{\dagger}} (t)~\sigma (t + \tau ) \rangle$, which is most relevant for the optical spectrum, as follows
\begin{equation}
\label{eq:populations2322}
\partial_\tau \mathbf{v} (t, t + \tau) = \mathbf{P} \langle {\sigma^{\dagger}} (t) \rangle - \mathbf{Q} \mathbf{v} (t, t + \tau),
\end{equation}
where the two-time correlators in $t$ and $t + \tau$, where $\tau$ is the delay time, are contained within the 3-vector
\begin{equation}
\label{eq:vmatrisx}
\mathbf{v} (t, t + \tau) =
\begin{pmatrix}
  \langle {\sigma^{\dagger}} (t)~\sigma (t + \tau ) \rangle \\
  \langle {\sigma^{\dagger}} (t)~\sigma^{\dagger} (t + \tau ) \rangle \\
  \langle {\sigma^{\dagger}} (t)~\sigma^{\dagger} (t + \tau ) \sigma (t + \tau ) \rangle
 \end{pmatrix}, 
 \end{equation}
and where the drive term $\mathbf{P}$ and the regression matrix $\mathbf{Q}$ are given in Eq.~\eqref{eqapp:umatrix}. In what follows, we consider zero detuning ($\Delta = 0$) for simplicity and we take the steady state limit $t \to \infty$, such that objects like $\langle {\sigma^{\dagger}} (t \to \infty) \rangle$ can be taken from Eq.~\eqref{eq:populationsd343434issi}.

The exact solution of Eq.~\eqref{eq:populations2322} is comprised of transient and steady state parts, as follows
\begin{equation}
\label{eq:solly1}
\mathbf{v} (t, t + \tau) = \sum_{\zeta=\mathrm{A}, \mathrm{B}, \mathrm{C}} a_{\zeta} \mathbf{v}_{\zeta}^E \mathrm{e}^{-\left(\mathrm{i} \omega_{\zeta} + \tfrac{\gamma_{\zeta}}{2} \right) \tau} 
 +
 \begin{pmatrix}
  c \\
  c^{\ast} \\
  n
 \end{pmatrix}
 \langle {\sigma^{\dagger}} (t) \rangle.
\end{equation}
In the first line of Eq.~\eqref{eq:solly1}, the $\zeta$-th complex eigenvalue of $-\mathbf{Q}$ is $\lambda_{\zeta}$, and it is associated with the eigenvector $\mathbf{v}_{\zeta}^E$. The complex eigenfrequencies $\lambda_{\zeta}$ may be decomposed as the damping decay rates $\gamma_{\zeta} = - 2 \mathrm{Re} \left( \lambda_{\zeta} \right)$ and the frequency shifts $\omega_{\zeta} = - \mathrm{Im} \left( \lambda_{\zeta} \right)$, producing the exponent in Eq.~\eqref{eq:solly1}. The constants $a_{\zeta}$ are to be found by imposing the steady state boundary conditions at zero delay time ($\tau = 0$), while the two quantities $c$ and $n$ are defined in Eq.~\eqref{eq:populationsd343434issi}.

With the knowledge of the two-time correlators relevant to the optical spectrum $s(\omega) = \langle \sigma^{\dagger} (\omega) \sigma (\omega) \rangle$, the spectrum normalized in the steady state $S( \omega )$ readily follows as~\cite{delValle2010,Camilo2016b, ValleLaussy2011, delVallePRL2010, Gardiner2014, ValleBook2010, Kavokin2007} 
\begin{equation}
\label{eq:pspec2}
 S(\omega)~=~S_{\mathrm{D}}~+ \sum_{\zeta = \mathrm{A}, \mathrm{B}, \mathrm{C}} S_{\zeta},
\end{equation}
where the objects inside of the summation in Eq.~\eqref{eq:pspec2} have been decomposed into the standard spectral lineshapes
\begin{equation}
\label{eq:pspec2343}
S_{\zeta} (\omega) = \frac{1}{\pi} \frac{\tfrac{\gamma_\zeta}{2}~L_\zeta - \left( \omega - \omega_\zeta \right)~K_\zeta}{ \left( \tfrac{\gamma_\zeta}{2} \right)^2 + \left( \omega - \omega_\zeta \right)^2  }.
\end{equation}
Here $L_\zeta$ and $K_\zeta$ are the real-valued weighting coefficients of the Lorentzian and dispersive parts respectively of Eq.~\eqref{eq:pspec2343}, as found from the formal solution of Eq.~\eqref{eq:solly1}. The so-called Rayleigh peak in Eq.~\eqref{eq:pspec2} is defined by
\begin{equation}
\label{eq:delta_peak}
S_{\mathrm{D}} (\omega) = L_{\mathrm{D}} \delta (\omega-\omega_0),
\end{equation}
where $\delta(x)$ is the Dirac delta function. Importantly, the complex eigenvalues $\lambda_{\zeta}$ of the dynamical matrix $-\mathbf{Q}$ appearing in Eq.~\eqref{eq:populations2322} define two regimes, defined by the critical driving strength
\begin{equation}
\label{eq:delta_crit}
\Omega_{\mathrm{c}} = \frac{\gamma}{8}.
\end{equation}
In what follows, we consider each case separately: firstly the supercritical case ($\Omega > \Omega_{\mathrm{c}}$), and secondly the subcritical case ($\Omega \le \Omega_{\mathrm{c}}$). 
\\


\begin{figure*}[tb]
 \includegraphics[width=1.0\linewidth]{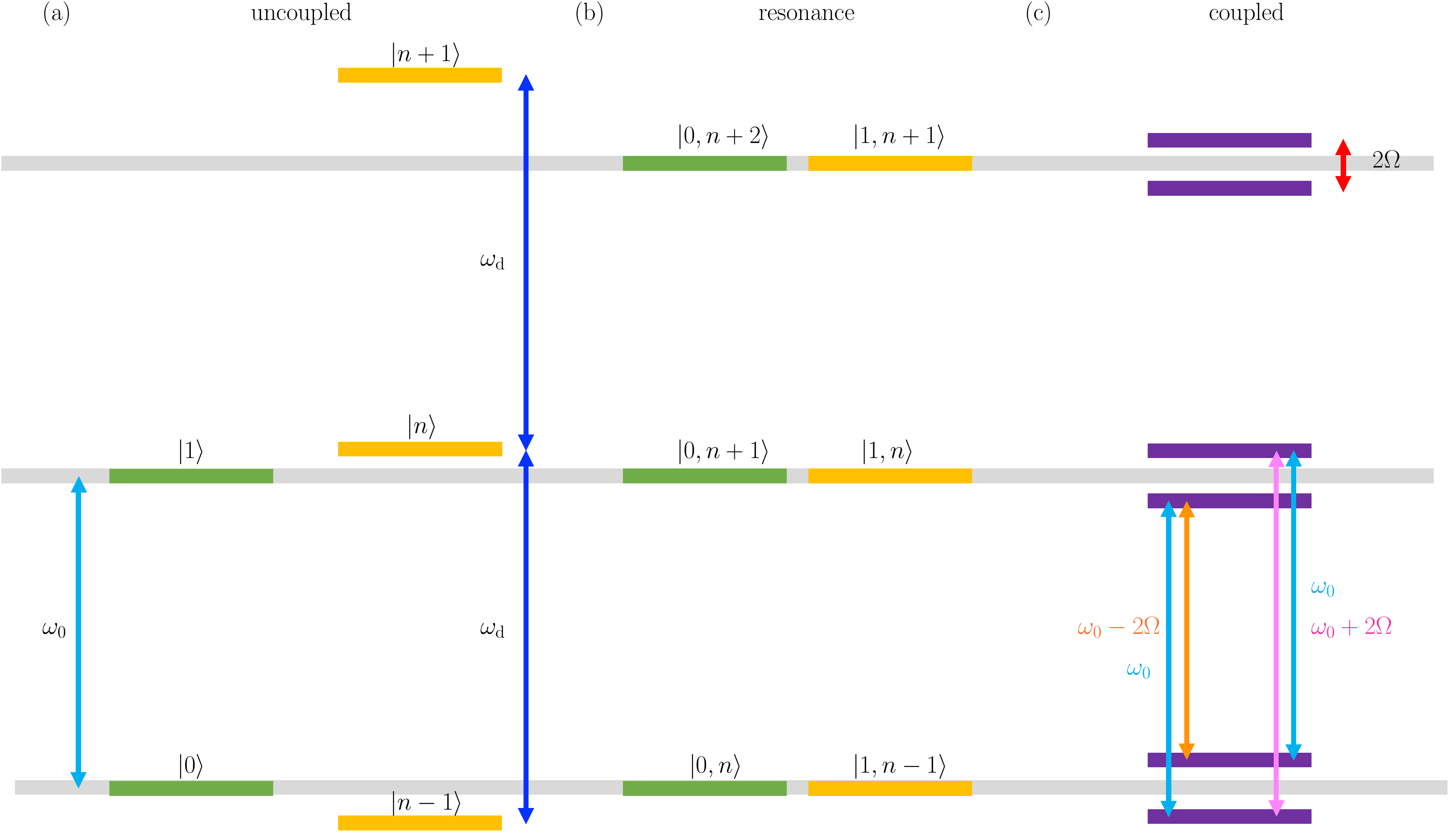}
 \caption{ \textbf{A sketch of the energy ladder giving rise to the Mollow triplet.}  Panel (a): the two states $| 0 \rangle$ and $| 1 \rangle$ of the two-level system are represented by green bars, and are separated by the frequency $\omega_0$ (cyan arrow). The infinite number of states $| n \rangle$ of the quantum harmonic oscillator (modelling the driving laser) are depicted by yellow bars, and the inter-level spacing $\omega_{\mathrm{d}}$ is shown (blue arrows). Panel (b): at resonance ($\omega_0 = \omega_{\mathrm{d}}$) and in the rotating frame of the driving laser, the combined system displays doubly-degenerate energy levels at $n \omega_0$ via the states $| 0, n \rangle$ and $| 1, n-1 \rangle$. Panel (c): the 2LS-laser coupling splits the doubly-degenerate levels, leading to a doublet spaced by $2 \Omega$ (red arrow). There are then four decay channels: two of which are associated with the transition frequency $\omega_0$ (cyan arrows), one with $\omega_0 + 2 \Omega$ (pink arrow) and one with $\omega_0 - 2 \Omega$ (orange arrow). These three distinct transition frequencies imply a spectral Mollow triplet.}
 \label{tran}
\end{figure*}

\noindent
\textbf{The supercritical regime and the Mollow triplet.} The three complex eigenvalues of $-\mathbf{Q}$ [cf. Eq.~\eqref{eqapp:umatrix}] may be decomposed in terms of their real and imaginary parts as $\tfrac{1}{2} \gamma_{\zeta} + \mathrm{i} \omega_{\zeta}$, leading to the exponent appearing in Eq.~\eqref{eq:solly1}. In the supercritical regime ($\Omega > \Omega_{\mathrm{c}}$), these eigenvalues are given by
\begin{subequations}
\label{eq:nmycoef}
 \begin{align}
 \tfrac{1}{2} \gamma_{\mathrm{A}} + \mathrm{i} \omega_{\mathrm{A}} &= \tfrac{1}{2} \gamma, \\
 \tfrac{1}{2} \gamma_{\mathrm{B}} + \mathrm{i} \omega_{\mathrm{B}} &= \tfrac{3}{4} \gamma + \mathrm{i} \Omega_{\mathrm{M}}, \\
 \tfrac{1}{2} \gamma_{\mathrm{C}} + \mathrm{i} \omega_{\mathrm{C}} &= \tfrac{3}{4} \gamma - \mathrm{i} \Omega_{\mathrm{M}},
  \end{align}
\end{subequations}
where we have introduced the Mollow frequency splitting $\Omega_{\mathrm{M}}$, defined by
\begin{equation}
\label{eq:MollowSplitting}
  \Omega_{\mathrm{M}} = \sqrt{ \left(2 \Omega \right)^2 - \left( \tfrac{\gamma}{4} \right)^2 }.
\end{equation}
The weighting coefficients $L_\zeta$ and $K_\zeta$ appearing in Eq.~\eqref{eq:pspec2343} provide the spectral weighting decomposition into real and imaginary parts as follows
\begin{widetext}
\begin{subequations}
\label{eq:nmycoef2}
 \begin{align}
 L_{\mathrm{A}}  &= \frac{1}{2}, \\
 L_{\mathrm{B}} + \mathrm{i} K_{\mathrm{B}}  &=  \frac{8 \Omega^2}{\gamma^2 + 8 \Omega^2 } \bigg\{ \frac{16 \Omega^2 - 2 \gamma^2}{ \left( 4 \Omega_{\mathrm{M}} \right)^2 + \gamma^2} + \mathrm{i} \frac{\gamma}{4 \Omega_{\mathrm{M}}} \frac{16 \Omega^2 - \gamma^2 + \left( 4 \Omega_{\mathrm{M}} \right)^2}{ \left( 4 \Omega_{\mathrm{M}} \right)^2 + \gamma^2} \bigg\}, \\
 L_{\mathrm{C}} + \mathrm{i} K_{\mathrm{C}}  &=  \frac{8 \Omega^2}{\gamma^2 + 8 \Omega^2 } \bigg\{ \frac{16 \Omega^2 - 2 \gamma^2}{ \left( 4 \Omega_{\mathrm{M}} \right)^2 + \gamma^2}  - \mathrm{i} \frac{\gamma}{4 \Omega_{\mathrm{M}}} \frac{16 \Omega^2 - \gamma^2 + \left( 4 \Omega_{\mathrm{M}} \right)^2}{ \left( 4 \Omega_{\mathrm{M}} \right)^2 + \gamma^2} \bigg\}, \\
 L_{\mathrm{D}} &=  \frac{\gamma^2}{\gamma^2 + 8 \Omega^2 }. 
  \end{align}
\end{subequations}
\end{widetext}
This analysis reveals the famous Mollow triplet~\cite{Mollow1969}, which can be interpreted via the progressive energy ladders sketched in Fig.~\ref{tran}. At either side of an unshifted central peak (denoted by $\mathrm{A}$), there are two satellite peaks (labelled by $\mathrm{B, C}$) and separated by the Mollow splitting $2 \Omega_{\mathrm{M}}$, along with a delta peak (denoted by $\mathrm{D}$) [cf. Eq.~\eqref{eq:delta_peak}]. The optical spectrum $S(\omega)$ follows by substituting Eqs.~\eqref{eq:nmycoef}~and~\eqref{eq:nmycoef2} into Eq.~\eqref{eq:pspec2}, leading to the exact expression [cf. Eq.~\eqref{eq:nmycdfsoef2sdsd2323}]
\begin{widetext}
\begin{align}
\label{eq:nmycoef2sdsd2323}
S( \omega ) &= \frac{\gamma^2}{\gamma^2 + 8 \Omega^2 } \delta (\omega -\omega_0) + \frac{1}{2 \pi} \frac{ \tfrac{\gamma}{2} }{ ( \tfrac{\gamma}{2} )^2 + \left( \omega -\omega_0 \right)^2 }  - \frac{\gamma}{\pi} \frac{\left( \omega -\omega_0 \right)^2 + \gamma^2 - 16 \Omega^2}{4 \left( \omega -\omega_0 \right)^4 + \left( \omega -\omega_0 \right)^2 \left( 5 \gamma^2 - 32 \Omega^2 \right) + \left( \gamma^2 + 8 \Omega^2 \right)^2 }.
\end{align}
\end{widetext}
We plot the spectrum $S(\omega)$ of Eq.~\eqref{eq:nmycoef2sdsd2323} in Fig.~\ref{single}~(b) for increasingly strong driving $\Omega$ (increasingly thick lines), and we have neglected the central Rayleigh delta peak. The Mollow triplet is most apparent for larger $\Omega$, where the three constituent peaks can be most easily resolved, see for example the case of $\Omega = \gamma$, as given by the thick purple line [cf. Eq.~\eqref{eq:MollowSplitting}]. This enchanting triplet structure was first observed in a series of pioneering experiments in the 1970's~\cite{Schuda1974, Wu1975, Hartig1976, Kimble1977}.
\\
 

\noindent
\textbf{The subcritical regime and the Mollow singlet.} In the subcritical regime ($\Omega \le \Omega_{\mathrm{c}}$) there are three real eigenvalues of $-\mathbf{Q}$, which lead to the decay rates [cf. Eq.~\eqref{eq:nmycoef}]
\begin{subequations}
 \begin{align}
\label{eq:nmycoef23}
 \gamma_{\mathrm{A}} &= \gamma, \\
 \gamma_{\mathrm{B}} &= \tfrac{3}{2} \gamma - 2 \gamma_{\mathrm{M}}, \\
 \gamma_{\mathrm{C}} &= \tfrac{3}{2} \gamma + 2 \gamma_{\mathrm{M}},
  \end{align}
\end{subequations}
where we have defined the Mollow decay rate $ \gamma_{\mathrm{M}}$ as [cf. Eq.~\eqref{eq:MollowSplitting}]
\begin{equation}
\label{eq:MollowSplitting2}
  \gamma_{\mathrm{M}} = \sqrt{ \left( \tfrac{\gamma}{4} \right)^2 - \left(2 \Omega \right)^2 }.
\end{equation}
The weighting coefficients in Eq.~\eqref{eq:pspec2343} are now wholly real numbers ($K_\zeta = 0$ for all $\zeta$), explicitly they are given by [cf. Eq.~\eqref{eq:nmycoef2}]
\begin{subequations}
 \begin{align}
\label{eq:nmycoef2544}
 L_{\mathrm{A}}  &= \tfrac{1}{2}, \\
 L_{\mathrm{B}}  &= \frac{ 8 \Omega^2 }{\gamma^2 + 8 \Omega^2 } \frac{\gamma^2 - 16 \Omega^2 + 4 \gamma \gamma_{\mathrm{M}} }{ \left( 4 \gamma_{\mathrm{M}} \right)^2 - 4 \gamma \gamma_{\mathrm{M}} }, \\
 L_{\mathrm{C}}  &= \frac{ 8 \Omega^2 }{\gamma^2 + 8 \Omega^2 } \frac{\gamma^2 - 16 \Omega^2 - 4 \gamma \gamma_{\mathrm{M}} }{ \left( 4 \gamma_{\mathrm{M}} \right)^2 + 4 \gamma \gamma_{\mathrm{M}} }, \\
 L_{\mathrm{D}}  &= \frac{\gamma^2}{\gamma^2 + 8 \Omega^2 },
  \end{align}
\end{subequations}
which together describe a single peak, unshifted in frequency. The complete spectrum $S( \omega )$ is given by Eq.~\eqref{eq:nmycoef2sdsd2323} and is plotted in Fig.~\ref{single}~(b). The singlet is most easily seen in the weak driving regime, see for example the case of $\Omega = \gamma/8$ (thin red line).
\\


\renewcommand{\theequation}{B \arabic{equation}}
\section{A pair of driven-dissipative two-level system}
\label{app:pop}


Here we show explicitly the calculations leading to the results presented in the main text for two coupled 2LSs. The mean populations and correlations are considered in Sec.~\ref{eqqy2}, and the corresponding optical spectrum is dealt with in Sec.~\ref{popcoh2}. Since we are dealing with a pair of few-level systems the Hilbert space is rather small, and there are a finite number of equations of motion for the moments.


\subsection{\label{eqqy2}Populations}

A generalization of the Hamiltonian given as Eq.~\eqref{eq:02} in the main text, in order to allow for independent coherent driving of both the first and the second 2LS respectively, can be captured by
\begin{align}
\label{eq:sdsdds02}
 \tilde{H} =& ~\Delta \left( \sigma_1^{\dagger} \sigma_1 + \sigma_2^{\dagger} \sigma_2 \right) + g \left( \mathrm{e}^{\mathrm{i} \theta} \sigma_1^{\dagger} \sigma_2 + \mathrm{e}^{- \mathrm{i} \theta} \sigma_2^{\dagger} \sigma_1 \right) \nonumber \\
  &+ \Omega_1 \left( \sigma_1 + \sigma_1^{\dagger} \right) + \Omega_2 \left( \sigma_2 + \sigma_2^{\dagger} \right),
\end{align}
where in the main text we effectively took $\Omega_1 = \Omega$ and $\Omega_2 = 0$. The quantum master equation of Eq.~\eqref{eq:master} from the main text, along with the above Eq.~\eqref{eq:sdsdds02}, leads to the following equation of motion for the mean values of the moments
\begin{equation}
\label{myeq:of_motion}
\partial_t \mathbf{u} = \mathbf{P} - \mathbf{M} \mathbf{u},
\end{equation}
where the 15-vector of correlators $\mathbf{u}$ and drive term $\mathbf{P}$ are given by
\begin{equation}
\label{eqerpp:umatrix}
\mathbf{u}  =
\begin{pmatrix}
  \mathbf{u}_1 \\
  \mathbf{u}_2 \\
  \mathbf{u}_3 \\
  \mathbf{u}_4
 \end{pmatrix},
 \quad \quad \quad 
\mathbf{P} = 
\begin{pmatrix}
  -\mathrm{i} \Omega_1  \\
   -\mathrm{i} \Omega_2 \\
   \mathrm{i} \Omega_1 \\
   \mathrm{i} \Omega_2 \\
  \text{\bf{0}}_{11}  
 \end{pmatrix},
 \end{equation}
where $\text{\bf{0}}_{n}$ is the zero matrix (of $n$-rows, and a single column). The vector of mean correlators $\mathbf{u}$ is further decomposed into 
\begin{equation}
\label{eqerpp:umatrix567u}
\mathbf{u}_1 =
\begin{pmatrix}
  \langle \sigma_1 \rangle \\
  \langle \sigma_2 \rangle \\
  \langle {\sigma_1^{\dagger}} \rangle \\
  \langle {\sigma_2^{\dagger}} \rangle
 \end{pmatrix},
 \quad  \quad
\mathbf{u}_2 =
\begin{pmatrix}
  \langle {\sigma_1^{\dagger}} \sigma_1 \rangle  \\
  \langle {\sigma_2^{\dagger}} \sigma_2 \rangle  \\
  \langle \sigma_1 \sigma_2  \rangle  \\
  \langle {\sigma_1^{\dagger}} {\sigma_2^{\dagger}} \rangle \\
  \langle {\sigma_1^{\dagger}} \sigma_2  \rangle  \\
  \langle \sigma_1 {\sigma_2^{\dagger}}  \rangle 
 \end{pmatrix},
 \end{equation}
 \begin{equation}
\label{eqerpp:umatrixdgfgd567u}
  \mathbf{u}_3 =
\begin{pmatrix}
  \langle {\sigma_1^{\dagger}} \sigma_1 \sigma_2 \rangle  \\
   \langle \sigma_1 {\sigma_2^{\dagger}} \sigma_2 \rangle  \\
  \langle {\sigma_1^{\dagger}} \sigma_1 {\sigma_2^{\dagger}} \rangle  \\
  \langle {\sigma_1^{\dagger}} {\sigma_2^{\dagger}} \sigma_2 \rangle 
 \end{pmatrix},
 \quad \quad  
 \mathbf{u}_4 = \langle {\sigma_1^{\dagger}} \sigma_1 {\sigma_2^{\dagger}} \sigma_2 \rangle.
 \end{equation}
The $15$-dimensional regression matrix $\mathbf{M}$ appearing inside Eq.~\eqref{myeq:of_motion} is given by
 \begin{equation}
\label{eqerpp:um678atrix56jju7u}
 \mathbf{M} =
\begin{pmatrix}
  \mathbf{M}_{11} && \mathbf{M}_{12} && \mathbf{M}_{13} && \text{\bf{0}}_{4, 1} \\
  \mathbf{M}_{21} && \mathbf{M}_{22} && \mathbf{M}_{23} && \mathbf{M}_{24} \\
  \text{\bf{0}}_{4, 4} && \mathbf{M}_{32} && \mathbf{M}_{33} && \mathbf{M}_{34} \\
  \text{\bf{0}}_{1, 4} && \text{\bf{0}}_{1, 6} && \mathbf{M}_{43} && 2 \gamma_0
 \end{pmatrix}, 
 \end{equation}
 where $\text{\bf{0}}_{n, m}$ is the zero matrix (of $n$-rows and $m$-columns). The eleven sub-matrices of $\mathbf{M}$ read
  \begin{equation}
\label{eqerpp:um678atrix5asd6jju7u}
 \mathbf{M}_{11} =
\begin{pmatrix}
  \tfrac{\gamma_0}{2} + \mathrm{i} \Delta & \tilde{g}_{+} & 0 & 0 \\
  \tilde{g}_{-}^{\ast} & \tfrac{\gamma_0}{2} + \mathrm{i} \Delta & 0 & 0 \\
 0 & 0 & \tfrac{\gamma_0}{2} - \mathrm{i} \Delta & \tilde{g}_{+}^{\ast} \\
  0 & 0 & \tilde{g}_{-} & \tfrac{\gamma_0}{2} - \mathrm{i} \Delta 
 \end{pmatrix},
  \end{equation}
    \begin{equation}
\label{eqerpp:um678atrix5zxcxczxcasd6jju7u}
  \mathbf{M}_{12} =
\begin{pmatrix}
   -2\mathrm{i} \Omega_1 & 0 & 0 & 0 & 0 & 0 \\
   0 &  -2\mathrm{i} \Omega_2 & 0 & 0 & 0 & 0 \\
  2\mathrm{i} \Omega_1 & 0 & 0 & 0 & 0 & 0 \\
 0 &  2\mathrm{i} \Omega_2 & 0 & 0 & 0 & 0  
 \end{pmatrix},
  \end{equation}
   \begin{equation}
\label{eqeadadsadsd6jju7u}
 \mathbf{M}_{13} =
\begin{pmatrix}
   -2\tilde{g}_{+} & 0 & 0 & 0 \\
    0 & -2\tilde{g}_{-}^{\ast} & 0 & 0 \\
     0 & 0 & -2\tilde{g}_{+}^{\ast} & 0 \\
      0 & 0 & 0 & -2\tilde{g}_{-}   
 \end{pmatrix},
  \end{equation}
     \begin{equation}
\label{eqeadadsadsd6jjdsfdfu7u}
  \mathbf{M}_{21} =
\begin{pmatrix}
   -\mathrm{i} \Omega_1 & 0 & \mathrm{i} \Omega_1 & 0 \\
   0 & -\mathrm{i} \Omega_2 & 0 & \mathrm{i} \Omega_2 \\
    \mathrm{i} \Omega_2 & \mathrm{i} \Omega_1 & 0 & 0 \\
     0 & 0 & -\mathrm{i} \Omega_2 & -\mathrm{i} \Omega_1 \\
      0 & -\mathrm{i} \Omega_1 & \mathrm{i} \Omega_2 & 0 \\
       -\mathrm{i} \Omega_2 & 0 & 0 & \mathrm{i} \Omega_1
 \end{pmatrix},
  \end{equation}
  \begin{equation}
\label{eqeadadsadvxdvyhhjju7u}
 \mathbf{M}_{22} =
\begin{pmatrix}
 \gamma_0 & 0 & 0 & 0 & \tilde{g}_{+} & \tilde{g}_{+}^{\ast} \\
 0 & \gamma_0 & 0 & 0 &\tilde{g}_{-} & \tilde{g}_{-}^{\ast} \\
  0 & 0 & \gamma_0 + 2 \mathrm{i} \Delta & 0 & 0 & 0 \\
   0 & 0 & 0 & \gamma_0 - 2 \mathrm{i} \Delta & 0 & 0 \\
    \tilde{g}_{-}^{\ast} & \tilde{g}_{+}^{\ast} & 0 & 0 & \gamma_0 & 0 \\
     \tilde{g}_{-} & \tilde{g}_{+} & 0 & 0 & 0 & \gamma_0 
 \end{pmatrix},
  \end{equation}
    \begin{equation}
\label{eqeadadsadvxdvyhhjju7u}
  \mathbf{M}_{23} =
\begin{pmatrix}
   0 & 0 & 0 & 0 \\
   0 & 0 & 0 & 0 \\
    -2\mathrm{i} \Omega_1 & -2\mathrm{i} \Omega_2 & 0 & 0 \\
     0 & 0 & 2\mathrm{i} \Omega_1 & 2\mathrm{i} \Omega_2 \\
      2\mathrm{i} \Omega_1 & 0 & 0 & -2\mathrm{i} \Omega_2 \\
       0 & 2\mathrm{i} \Omega_2 & -2\mathrm{i} \Omega_1 & 0
 \end{pmatrix},
  \end{equation}
    \begin{equation}
\label{xxeqaxsxaadsadvxdvyhhjju7u}
 \mathbf{M}_{24} =
\begin{pmatrix}
\text{\bf{0}}_{4} \\
 -2\gamma \mathrm{e}^{-\mathrm{i} \phi} \\
 -2\gamma \mathrm{e}^{\mathrm{i} \phi}   
 \end{pmatrix},
  \end{equation}
      \begin{equation}
\label{xxeqaxsxaadsaasdsadaddvxdvyhhjju7u}
 \mathbf{M}_{32} =
\begin{pmatrix}
   \mathrm{i} \Omega_2 & 0 & -\mathrm{i} \Omega_1 & 0 & \mathrm{i} \Omega_1 & 0 \\
   0 & \mathrm{i} \Omega_1 & -\mathrm{i} \Omega_2 & 0 & 0 & \mathrm{i} \Omega_2 \\
   -\mathrm{i} \Omega_2 & 0 & 0 & \mathrm{i} \Omega_1 & 0 & -\mathrm{i} \Omega_1 \\
   0 & -\mathrm{i} \Omega_1 & 0 & \mathrm{i} \Omega_2 & -\mathrm{i} \Omega_2 & 0 
 \end{pmatrix},
  \end{equation}
  \begin{equation}
\label{xxeqaxsadsxaadsadvxdvyhhjju7u}
 \mathbf{M}_{33} =
\begin{pmatrix}
  \tfrac{3\gamma_0}{2} + \mathrm{i} \Delta & \tilde{g}_{+}^\ast & 0 & 0 \\
  \tilde{g}_{-} & \tfrac{3\gamma_0}{2} + \mathrm{i} \Delta & 0 & 0 \\
  0 & 0 & \tfrac{3\gamma_0}{2} - \mathrm{i} \Delta & \tilde{g}_{+} \\
  0 & 0 & \tilde{g}_{-}^\ast & \tfrac{3\gamma_0}{2} - \mathrm{i} \Delta \\
 \end{pmatrix},
   \end{equation}
     \begin{equation}
\label{xxeqaxsadsxaadsdfdfdadvxdvyhhjju7u}
  \mathbf{M}_{34} =
\begin{pmatrix}
   -2\mathrm{i} \Omega_2 \\
   -2\mathrm{i} \Omega_1 \\
   2\mathrm{i} \Omega_2 \\
   2\mathrm{i} \Omega_1 
 \end{pmatrix},
   \end{equation}
  \begin{equation}
\label{eq:gqdfdfduantity}
 \mathbf{M}_{43} =
\begin{pmatrix}
   -\mathrm{i} \Omega_2 ~~ & -\mathrm{i} \Omega_1 ~~~ & \mathrm{i} \Omega_2 ~~~ & \mathrm{i} \Omega_1 \\ 
 \end{pmatrix}.
\end{equation}
In the above sub-matrices $\mathbf{M}_{nm}$, we have introduced the generalized coupling constants $\tilde{g}_{\pm}$, defined as
\begin{equation}
\label{eq:gquantity}
 \tilde{g}_{\pm} = \pm \mathrm{i} g \mathrm{e}^{\mathrm{i} \theta} + \tfrac{1}{2} \gamma \mathrm{e}^{\mathrm{i} \phi},
\end{equation}
which account for the competition between the coherent ($g \mathrm{e}^{\mathrm{i} \theta}$) and dissipative ($\gamma \mathrm{e}^{\mathrm{i} \phi}$) coupling within the coupled 2LS system. Notably, one can see that when the unidirectional conditions of Eq.~\eqref{eq:conditions} in the main text are fulfilled, one of either $\tilde{g}_{+}$ or $\tilde{g}_{-}$ vanishes, whilst the other remains finite. This is a basic fingerprint hinting that one-way coupling has arisen in the system.  

In the steady state ($\mathrm{ss}$), where $t \to \infty$ and $\partial_t \mathbf{u} = 0$ on the left-hand-side of Eq.~\eqref{myeq:of_motion}, we are interested in several time-independent quantities. Namely, the probabilities of having the first and second 2LS respectively excited, $n_{1} = \langle \sigma_{1}^{\dagger} \sigma_{1} \rangle_{\mathrm{ss}}$ and $n_{2} = \langle \sigma_{2}^{\dagger} \sigma_{2} \rangle_{\mathrm{ss}}$, and the joint probability that both 2LSs are excited, $n_{\mathrm{X}} = \langle \sigma_1^{\dagger} \sigma_1 \sigma_2^{\dagger} \sigma_2 \rangle_{\mathrm{ss}}$. We also have access to the probabilities of having only 2LS-1 excited $\rho_{1, 0} = n_{1} - n_{\mathrm{X}}$, only 2LS-2 excited $\rho_{0, 1} = n_{2} - n_{\mathrm{X}}$, the population of the ground state $\rho_{0, 0} = 1 + n_{\mathrm{X}} - n_1 - n_2$, and the population of the doubly-excited state $\rho_{1, 1} = n_{\mathrm{X}}$. Of course, unitarity is preserved and $\rho_{0, 0} + \rho_{1, 0} + \rho_{0, 1} + \rho_{1, 1} = 1$. The steady populations across the coupling landscape are considered in detail in the main text.


\subsection{\label{popcoh2}Spectrum}

\begin{figure*}[tb]
 \includegraphics[width=\linewidth]{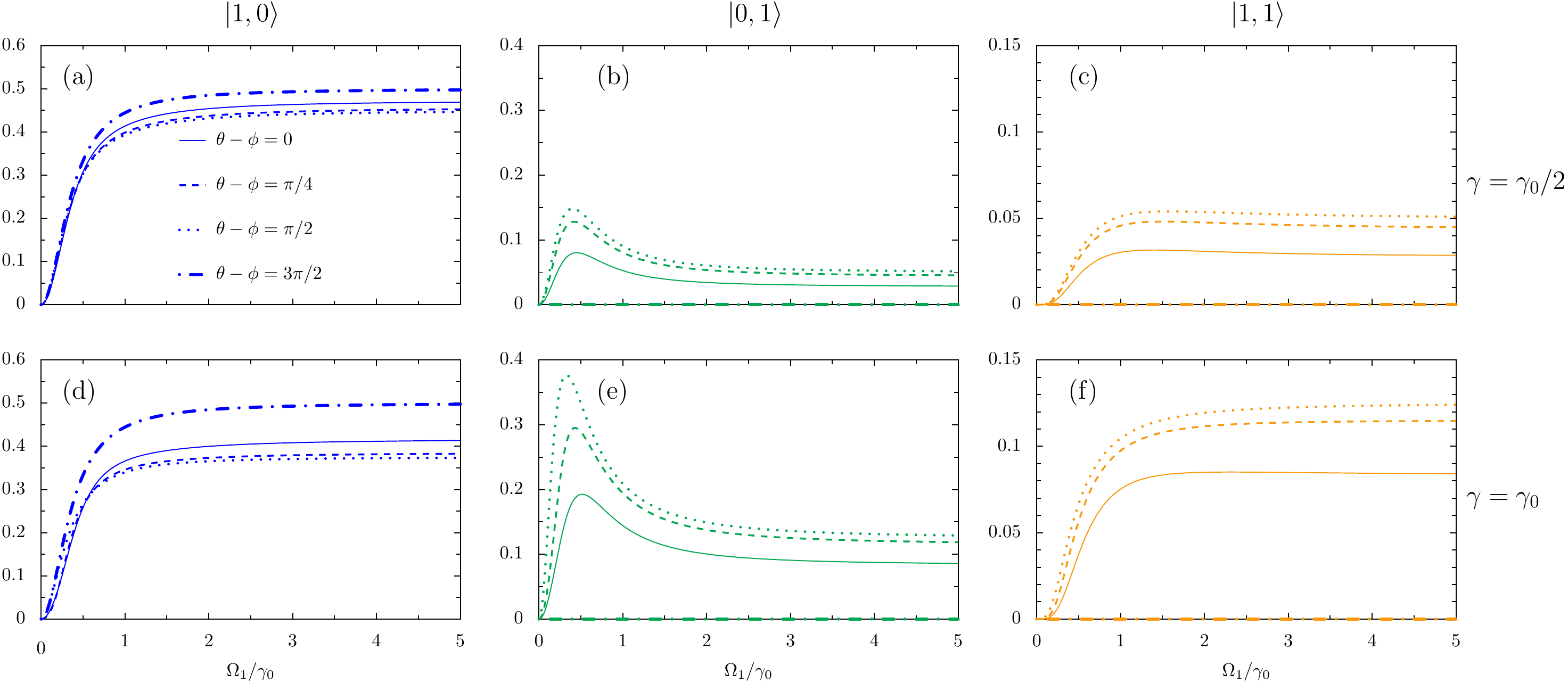}
 \caption{ \textbf{Populations in the asymmetric coupling regime.} Mean populations as a function of the drive amplitude $\Omega$, in units of the damping rate $\gamma_0$. The unidirectional magnitude condition $g = \gamma/2$ is fixed throughout. The relative phases $\theta-\phi = \{0, \pi/4, \pi/2, 3\pi/2 \}$ are denoted by solid, dashed, dotted, and dash-dotted lines respectively. Top row: the dissipative coupling strength $\gamma = \gamma_0/2$. Bottom row: $\gamma = \gamma_0$. The labeling of the mean population of the state $\ket{n, m}$ is displayed above the top row of panels. In the figure, the parameters $\Delta = \Omega_2 = 0$. }
 \label{popgen}
\end{figure*}

The quantum master equation of Eq.~\eqref{eqapp:master} from the main text, combined with the quantum regression formula, gives rise to the following two-time ($t$ and $t+\tau$, with $\tau \ge 0$ being the delay time) equation of motion [cf. Eq.~\eqref{myeq:of_motion}]
\begin{equation}
\label{myeq:of_motisdsdsdsdon}
\partial_\tau \mathbf{v} (t, t + \tau) = \mathbf{P} \langle \sigma_1^\dagger (t) \rangle - \mathbf{M} \mathbf{v} (t, t + \tau),
\end{equation}
where the column vector $\mathbf{P}$ is defined in Eq.~\eqref{eqerpp:umatrix}, and the $15 \times 15$ regression matrix $\mathbf{M}$ is given by Eq.~\eqref{eqerpp:um678atrix56jju7u}. The 15-vector of two-time correlators $\mathbf{v}$ is decomposed into [cf. Eq.~\eqref{eqerpp:umatrix}]
\begin{equation}
\label{eqerpp:umatsdsdsrix}
\mathbf{v} (t, t + \tau) =
\begin{pmatrix}
  \mathbf{v}_1 \\
  \mathbf{v}_2 \\
  \mathbf{v}_3 \\
  \mathbf{v}_4
 \end{pmatrix},
 \end{equation}
where the four component sub-vectors $\mathbf{v}_n$ read
\begin{equation}
\label{eqerpp:umatrsdsdsix567u}
\mathbf{v}_1 =
\begin{pmatrix}
  \langle \sigma_1^\dagger (t) \sigma_1 (t, t + \tau) \rangle \\
  \langle \sigma_1^\dagger (t) \sigma_2 (t, t + \tau) \rangle \\
  \langle \sigma_1^\dagger (t) {\sigma_1^{\dagger}} (t, t + \tau) \rangle \\
  \langle \sigma_1^\dagger (t) {\sigma_2^{\dagger}} (t, t + \tau) \rangle
 \end{pmatrix},
 \end{equation}
 \begin{equation}
\label{eqerpp:umatryiuyuisdsdsix567u}
 \mathbf{v}_2 =
\begin{pmatrix}
  \langle \sigma_1^\dagger (t) \sigma_1^{\dagger} (t, t + \tau) \sigma_1 (t, t + \tau) \rangle  \\
  \langle \sigma_1^\dagger (t) \sigma_2^{\dagger} (t, t + \tau) \sigma_2 (t, t + \tau) \rangle  \\
  \langle \sigma_1^\dagger (t) \sigma_1 (t, t + \tau) \sigma_2  (t, t + \tau) \rangle  \\
  \langle \sigma_1^\dagger (t) \sigma_1^{\dagger} (t, t + \tau) \sigma_2^{\dagger} (t, t + \tau) \rangle \\
  \langle \sigma_1^\dagger (t) \sigma_1^{\dagger} (t, t + \tau) \sigma_2  (t, t + \tau) \rangle  \\
  \langle \sigma_1^\dagger (t) \sigma_1(t, t + \tau) \sigma_2^{\dagger} (t, t + \tau) \rangle 
 \end{pmatrix},
 \end{equation}
 \begin{equation}
\label{eqerpp:umatrsdcdweweewsdsix567u}
  \mathbf{v}_3 =
\begin{pmatrix}
  \langle \sigma_1^\dagger (t) \sigma_1^{\dagger} (t, t + \tau) \sigma_1 (t, t + \tau) \sigma_2 (t, t + \tau) \rangle  \\
   \langle \sigma_1^\dagger (t) \sigma_1 (t, t + \tau)  \sigma_2^{\dagger} (t, t + \tau) \sigma_2 (t, t + \tau) \rangle  \\
  \langle \sigma_1^\dagger (t) \sigma_1^{\dagger} (t, t + \tau) \sigma_1 (t, t + \tau) \sigma_2^{\dagger} (t, t + \tau) \rangle  \\
  \langle \sigma_1^\dagger (t) \sigma_1^{\dagger} (t, t + \tau) \sigma_2^{\dagger} (t, t + \tau) \sigma_2 (t, t + \tau) \rangle 
 \end{pmatrix},
 \end{equation}
  \begin{equation}
\label{eqerpp:umsdfsdfdfsatsdssfddsdrix56jju7u}
 \mathbf{v}_4 = \langle \sigma_1^\dagger (t) \sigma_1^{\dagger} (t, t + \tau) \sigma_1 (t, t + \tau) \sigma_2^{\dagger} (t, t + \tau) \sigma_2 (t, t + \tau) \rangle.
 \end{equation}
The two-time correlator $\langle \sigma_1^\dagger (t) \sigma_1 (t, t + \tau) \rangle$ arising from the solution of Eq.~\eqref{myeq:of_motisdsdsdsdon} is intrinsically linked to the optical spectrum of the first 2LS, $s_1(\omega) = \langle \sigma_1^{\dagger} (\omega) \sigma_1  (\omega) \rangle$, via a Fourier transform and an application of the Weiner-Khinchin theorem. This procedure leads to the normalized spectrum $S_1(\omega)$, in exactly the same way as for the spectrum $S(\omega)$ of a single 2LS as was discussed in Appendix~\ref{app:single_2LS}, or indeed Refs.~\cite{delValle2010, ValleLaussy2011, delVallePRL2010, ValleBook2010, Kavokin2007}.


\renewcommand{\theequation}{C \arabic{equation}}
\section{Asymmetric coupling}
\label{gengen}


Here we focus on the most general coupling regime where there are essentially no restrictions on the system parameters, such that the coupling is typically asymmetric. Within this regime, we discuss the mean populations in Sec.~\ref{genpop}, the second-order degree of coherence in Sec.~\ref{gencorr} and the optical spectrum in Sec.~\ref{genspec}.


\subsection{\label{genpop}Populations}

\begin{figure*}[tb]
 \includegraphics[width=\linewidth]{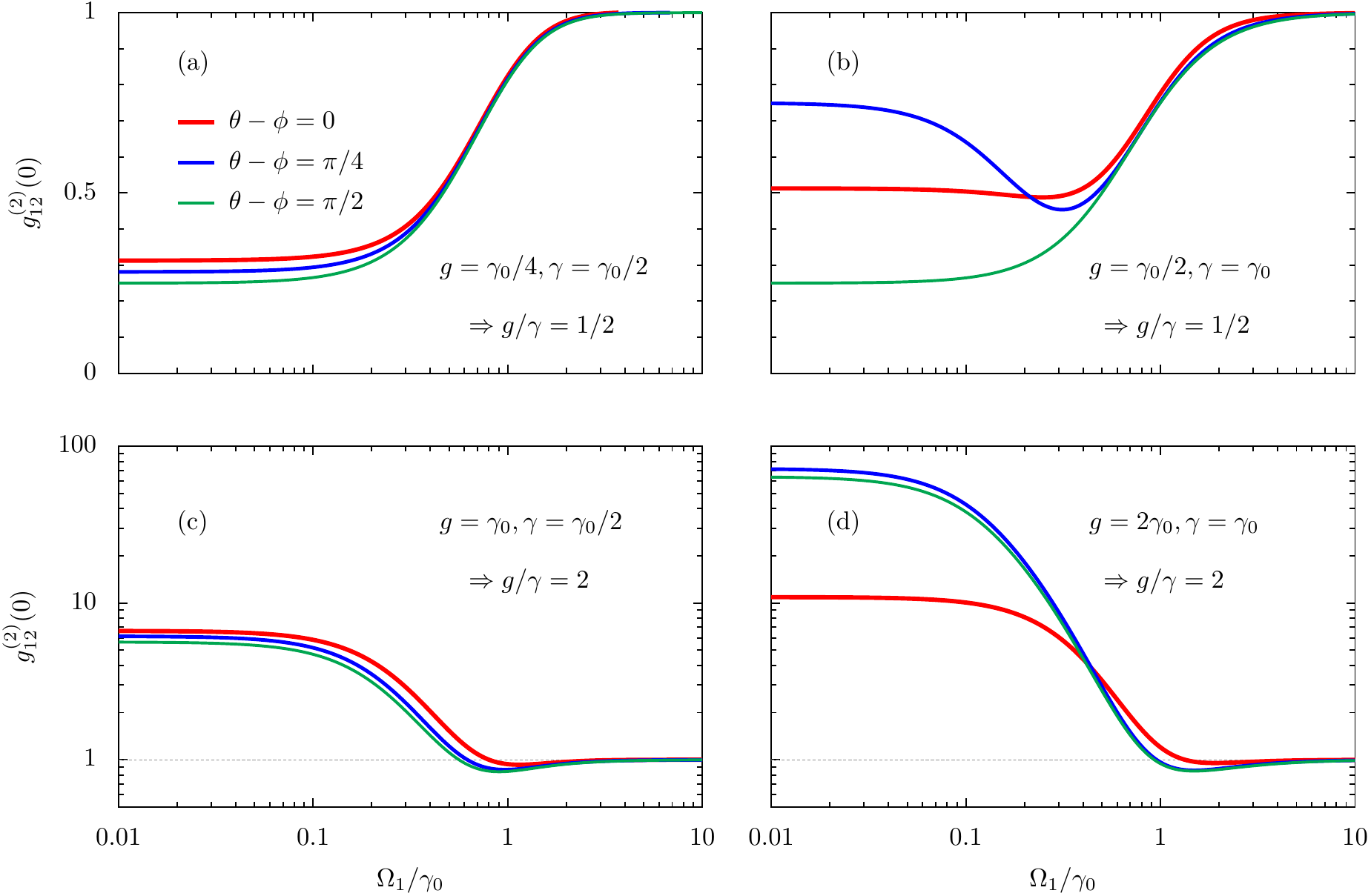}
 \caption{ \textbf{Correlations in the asymmetric coupling regime.} Cross-correlator $g_{12}^{(2)} (0)$, as a function of the drive amplitude $\Omega$, in units of the damping rate $\gamma_0$. The relative phases $\theta-\phi = \{0, \pi/4, \pi/2 \}$ are denoted by red, blue and green lines respectively. Upper panels: the relative coupling strength is fixed at the unidirectional magnitude condition of $g/\gamma = 1/2$, where in panel (a) $\gamma = \gamma_0/2$ and in panel (b) $\gamma = \gamma_0$. Lower panels: the relative coupling strength is fixed at $g/\gamma = 2$, where in panel (c) $\gamma = \gamma_0/2$ and in panel (d) $\gamma = \gamma_0$. Dashed gray lines: guides for the eye at $g_{12}^{(2)} (0) = 1$. }
 \label{corrgen}
\end{figure*}

We plot in Fig.~\ref{popgen} the mean population in the asymmetric coupling case, where the unidirectional magnitude condition $g = \gamma/2$ is fixed throughout. The relative phases $\theta-\phi = \{0, \pi/4, \pi/2, 3\pi/2 \}$ are denoted by solid, dashed, dotted and dash-dotted lines respectively in the figure. Going from left to right across the rows, the mean populations of the states $\ket{1, 0}, \ket{0, 1}, \ket{1, 1}$ are displayed in blue, green and orange respectively. In the top panels the dissipative coupling strength $\gamma = \gamma_0/2$, and in the lower panels it is increased to its maximal value of $\gamma = \gamma_0$. In panels (a) and (d) of Fig.~\ref{popgen}, the highest populations of the state $\ket{1, 0}$ are associated with the dash-dotted lines, representing $\theta-\phi = 3\pi/2$ and therefore unidirectional coupling from 2LS-2 to 2LS-1. Conversely, the lowest populations of $\ket{1, 0}$ are found when $\theta-\phi = \pi/2$ (dotted lines), since there is no backaction coming from 2LS-2. Similarly, in panels (b) and (e) the highest populations of the state $\ket{0, 1}$ correspond to the dotted lines, representing $\theta-\phi = \pi/2$, and thus forbidding any backaction into the second 2LS. Meanwhile, the dash-dotted lines in these panels equate to zero population, due to the one-way nature of the coupling. Panels (c) and (f) display the analogous results for the doubly excited state $\ket{1, 1}$ for completeness. Clearly, the variation in relative phase $\theta-\phi$ opens up the freedom to significantly modulate the steady state populations of all of the states, which is most noticeable in the lower row of panels (d, e, f) due to the stronger coupling.


\subsection{\label{gencorr}Correlations}

We plot in Fig.~\ref{corrgen} the cross-correlator $g_{12}^{(2)} (0)$ as a function of the drive amplitude $\Omega$, where the relative phases $\theta-\phi = \{0, \pi/4, \pi/2 \}$ are denoted by red, blue and green lines respectively. In the upper panels, the relative coupling strength is fixed at the unidirectional magnitude condition of $g/\gamma = 1/2$, where in panel (a) $\gamma = \gamma_0/2$ and in panel (b) $\gamma = \gamma_0$. Both panels (a) and (b) showcase antibunching $g_{12}^{(2)} (0) < 1$, and it is noticeable that with weaker coupling in panel (a) the effect of the relative phase $\theta-\phi$ is much reduced. With stronger coupling in panel (b), the relative phases becomes much more important, and especially determining with weak coupling $\Omega \ll \gamma_0$. In the lower panels of Fig.~\ref{corrgen}, the relative coupling strength is fixed at the higher ratio $g/\gamma = 2$, so unidirectional coupling is no longer possible. In panel (c) the dissipative coupling is moderate ($\gamma = \gamma_0/2$), and in panel (d) it is maximal ($\gamma = \gamma_0$). Consequentially, now bunching $g_{12}^{(2)} (0) > 1$ is primarily displayed with weak driving $\Omega \ll \gamma_0$. The most extreme bunching is showcased in panel (d) due to the stronger coupling, especially for nonreciprocal phases (thinner blue and green lines). The figure highlights the tremendous variety in the system which is opened up due to the interplay of coherent and incoherent coupling.


\subsection{\label{genspec}Spectrum}

\begin{figure*}[tb]
 \includegraphics[width=\linewidth]{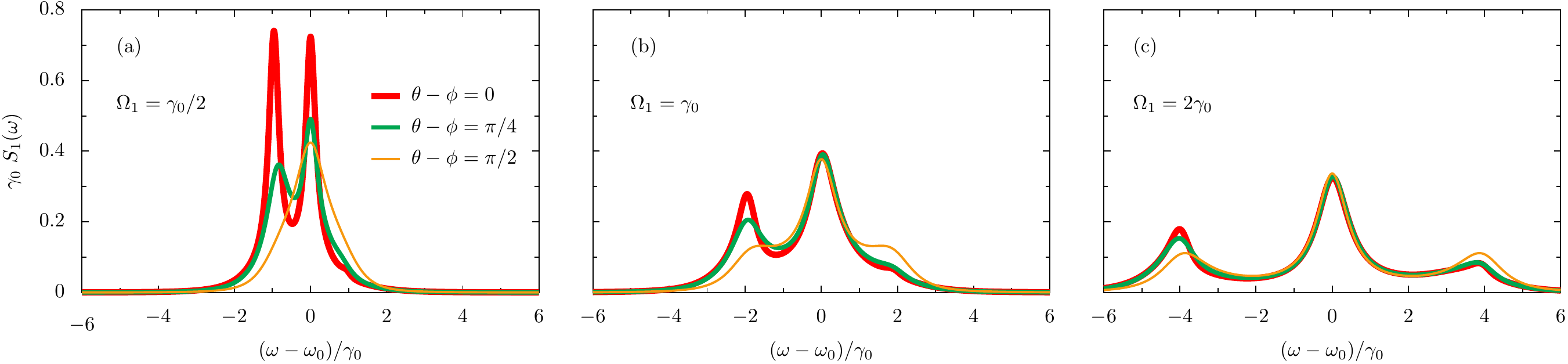}
 \caption{ \textbf{Spectra in the asymmetric coupling regime.} Optical spectrum $S_1(\omega)$ of the first 2LS in the asymmetric coupling regime, in units of the inverse damping rate $\gamma_0^{-1}$. In each panel, increasingly large relative phases $\theta - \phi = \{ 0, \pi/4, \pi/2\}$ are denoted by increasingly thick colored lines. Panels (a): the driving strength $\Omega = \gamma_0/2$. Panels (b): $\Omega = \gamma_0$. Panels (c): $\Omega = 2\gamma_0$. In the figure, the unidirectional magnitude condition $g = \gamma/2$ is observed, and the dissipative coupling strength is maximal $\gamma = \gamma_0$. }
 \label{specgen}
\end{figure*}

In Fig.~\ref{specgen}, we show the optical spectrum $S_1(\omega)$ of the first 2LS in the asymmetric coupling regime. In each panel, increasingly large relative phases $\theta - \phi = \{ 0, \pi/4, \pi/2\}$ are denoted by increasingly thick colored lines, and we fix $g = \gamma/2$ and $\gamma = \gamma_0$. In panel (a), the driving strength $\Omega = \gamma_0/2$, such that for the unidirectional case with $\theta - \phi = \pi/2$ (thin orange line) the spectrum is a standard singlet. Remarkably, for the other relative phases (thicker green and red lines) a striking asymmetric doublet appears due to the asymmetric coupling. In Fig.~\ref{specgen}~(b), the driving strength is increased to $\Omega = \gamma_0$, so that the spectrum is a symmetric Mollow triplet in the unidirectional coupling regime (thin orange line). Away from this special case, an asymmetry again appears (thicker green and red lines) so that the spectrum almost approaches a doublet form, with only a small third peak. The driving strength is further increased to $\Omega = 2\gamma_0$ in Fig.~\ref{specgen}~(c). The spectrum is now a widely separated symmetric triplet in the unidirectional coupling regime (thin orange line). The other phases (thicker green and red lines) lead to an skewed triplet spectrum, which is a hallmark of asymmetric coupling in the pair of 2LSs. The figure represents how the modulation of the phase $\theta - \phi$ can lead to significant reconstructions of the spectrum, especially with regard to asymmetries and additional sidebands not previously possible.

\end{document}